\documentclass{article}
\usepackage[a4paper, left=2cm, right=2cm, top=2cm]{geometry}
\usepackage[utf8]{inputenc}
\usepackage[ruled]{algorithm2e}

\SetAlFnt{\small}
\SetAlCapFnt{\small}
\SetAlCapNameFnt{\small}
\SetAlCapHSkip{0pt}
\IncMargin{-\parindent}
\usepackage{graphics}      
\usepackage[T1]{fontenc}   
\usepackage{pdfpages}
\usepackage{txfonts}
\usepackage{mathptmx}
\usepackage[pdflang={en-US},pdftex]{hyperref}
\usepackage{color}
\usepackage{booktabs}
\usepackage{textcomp}
\usepackage[english]{babel}
\usepackage{hyphenat}
\usepackage{authblk}
\usepackage{multicol}

\title{That Depends - Assessing User Perceptions of Authentication Schemes across Contexts of Use}

\author[1]{Verena Zimmermann}
\author[2]{Paul Gerber}
\author[2]{Alina Stöver}
\affil[1]{ETH Zürich \footnote{Verena Zimmermann was previously affiliated with Technische Universität Darmstadt where this research work has been conducted.}}
\affil[2]{Technische Universität Darmstadt}
\date{}

\begin{document}

\maketitle

\begin{abstract}
Choosing authentication schemes for a specific purpose is challenging for service providers, developers, and researchers. Previous ratings of technical and objective aspects showed that available schemes all have strengths and limitations. Yet, the security of authentication also relies on user perceptions which affect acceptance and user behaviour and can deviate from technical aspects. To shine light on the issue and support researchers, developers, and service-providers confronted with authentication choice, we conducted an in-depth analysis of user perceptions of the password, fingerprint, and a smartphone-based scheme in an online study with 201 participants. As authentication is a secondary task that needs to be evaluated in the context of authentication purpose, we also compared perceptions across four contexts of use with varying sensitivity levels: email accounts, online banking, social networks, and smart homes. The results revealed how perceptions of usability, security, privacy, trust, effort, and qualitative features of the schemes are related to user preferences. The results increase awareness for the influence of subjective perceptions and have practical implications for decision-makers. They can inform a) the choice between several adequate schemes, b) the authentication design to reduce concerns or security-related misconceptions, and c) the development of context-dependent authentication. 
\end{abstract}



\section{Introduction}
Authentication schemes serve to protect personal and valuable data from unauthorised access. They are commonly divided into knowledge-based (something you know), biometric (something you are), and token-based schemes (something you have) \cite{grassi2017digital}. Yet, selecting a suitable authentication scheme amongst the plethora of available schemes is a challenging task. A previously conducted rating of technical and objective aspects 
\cite{bonneau2012quest,renaud2014access,mayer2016supporting,zimmermann2019keep} 
revealed that none of the tested schemes outperformed the others in all regards. Instead, each scheme demonstrated strengths in certain areas and weaknesses in others. For example, the ubiquitous password scored well in terms of deployability as compared to biometric schemes such as the fingerprint \cite{bonneau2012quest}. Likewise, biometrics received higher scores in terms of certain security features compared to passwords \cite{bonneau2012quest}. 

Aiming to facilitate choice, previous research has often focused on comparing and improving objective and technical aspects of existing and newly developed schemes, such as the resistance against certain attacks \cite{he2016efficient,o2003comparing,wang2007cryptanalysis,zhao2007s3pas} or the efficiency or accuracy of algorithms \cite{he2016efficient,li2010efficient,wang2007cryptanalysis}.


However, despite these efforts, no clear ``silver bullet'' emerged from the vast selection, neither in terms of single-factor authentication \cite{bonneau2012quest,herley2011research,zimmermann2019keep} nor in terms of the ideal combination of schemes in multi-factor authentication (MFA) \cite{velasquez2019multifactor,wang2016request}. 
This leaves researchers, developers, and service providers with the difficult decision of choosing suitable authentication schemes or combinations thereof for their service or database.  

Two aspects that have been studied less frequently as compared to technical aspects might shine light on this seemingly intractable issue of authentication choice: First, the subjective user perceptions of authentication schemes, and second, the context of authentication. This research fills the existing gap by comprehensively studying user perceptions of authentication schemes across contexts of use with a large sample of \textit{N}=201 users.

\subsection{Subjective User Perceptions}

Subjective user perceptions affect the way security technologies are used as well as the extent to which they are adopted \cite{garfinkel2014usable}. 
Yet, research shows that users have difficulties assessing security properties and that user perceptions can deviate from technical security features \cite{ur2016users,ur2015added,zimmermann2019password}.
This can negatively affect security: Users might refrain from using an actually secure technology if the perceived security is too low \cite{huang2011factors}. Similarly, users might engage in insecure practices if users perceive security to be high even if it is not. Considering user perceptions is therefore not only relevant in terms of acceptance and adoption, but also in terms of authentication security.

\subsection{Authentication Context}

Similar to many security-related tasks, authentication is a secondary task \cite{whitten1999johnny}. Users aim to log in to use a certain service and to access certain kinds of data, not merely to log in. However, the types of services and related data are of different value and sensitivity. Whereas financial data such as the credit card number, or health information such as the medical history, are rated as highly sensitive, revealing the favourite TV show, the email address, or basic demographics is perceived as less critical \cite{ackerman1999privacy,milne2017information,yang2009influence}. Research indicates that people are inclined to invest more in security, e.g., by using MFA or more secure passwords, for sensitive accounts \cite{abbott2020mandatory,sun2008users,stobert2018password}. Cumbersome authentication for short-term access to non-sensitive data can be perceived as unnecessary \cite{harbach2014sa,sun2011more} and potentially leads to frustration or rejection by users.


It is thus important to consider these aspects for identifying a suitable trade-off concerning authentication choice for a certain context or type of data, respectively. In line with that, researchers also suggest undertaking further efforts for selecting and designing content- and context-dependent authentication mechanisms \cite{harbach2014sa,hayashi2013casa,hayashi2012goldilocks}. 

\subsection{Research Question and Hypotheses}

To shine light on the users' perceptions of different authentication schemes, this research quantitatively and qualitatively explores differences across different types of authentication schemes in a within-subject study with \textit{N}=201 participants. To do so, we chose representatives for each of the three categories due to their popularity and previous research results: password authentication (something you know), fingerprint authentication (something you are), and smartphone-based authentication (something you have). To address the service and the type of data that is accessed, we also explore differences across different contexts of use.

This research aims to answer the following research question: \textit{How do user perceptions differ across authentication schemes and contexts of use?}
To do so, several hypotheses are tested:

\textbf{$H_1$:} There are differences in user perceptions across different authentication schemes.  

\textbf{$H_2$:} There are differences in user perceptions of an authentication scheme across different contexts of use.

\textbf{$H_3$:} There are differences in user perceptions of different authentication schemes across different contexts of use. 

\textbf{$H_4$:} User preference in a certain context of use is influenced by the context-specific user perceptions in terms of the scheme, as well as general perceptions of the scheme and the context. 

\subsection{Findings \& Contribution}

The contribution of this research is threefold: 

\textbf{Contribution 1: } This research provides an in-depth analysis of user perceptions of authentication schemes across different contexts of use. By studying both aspects in unison and with a large and heterogeneous sample, it extends previous comparisons of authentications schemes \cite{abbott2020mandatory,furnell2007public,gerber2017security,heckle2007perception,jones2007towards,ruoti2015authentication}.

In terms of the first contribution, the study results revealed that the type of authentication scheme, as well as the context of use individually, impact user perceptions of authentication. Furthermore, the interaction of scheme and context of use had a significant influence on the users' perceptions. While the preference, intention to use, and effort-benefit ratio ratings differed with the scheme and context, the privacy and security perceptions seemed to depend on the scheme and remained stable across contexts of use. The results confirm $H_1$, $H_2$, and $H_3$.

Overall, password authentication was highly preferred despite the cognitive load it poses for users and fingerprint authentication being perceived as more secure. Preference ratings in terms of fingerprint authentication were ambiguous, 
perhaps due to privacy concerns with sharing biometric information for authentication purposes. 

\textbf{Contribution 2:} Beyond a detailed comparison of differences in user perceptions, this research explores how different factors impact user preferences. Unlike many previous studies, this research explores the influence of numerous potential influencing factors at the same time, so that they can be compared against each other. 

It was found that perceived security and effort-benefit ratio, as well as the general perceived usability of a scheme predicted user preference across different schemes and contexts. Therefore, $H_4$ can be partially confirmed. Moreover, while there were some indications for the influence of familiarity on user perceptions, it did not appear as a significant predictor for preference, potentially because of the type of measurement deployed. 

\textbf{Contribution 3:} The findings lend guidance for decision-makers in terms of the perceptions and contextual factors to consider when choosing, designing, and implementing user-centred authentication. They can be used by decision-makers to break a tie between several suitable schemes, to address concerns and potentially security-critical misconceptions in the system design, or to inform the design of context-dependent authentication.

For example, decision-makers might decide to implement fingerprint authentication for a context with highly sensitive data, such as a smartphone banking application, based on the users' high preference and perceived effort-benefit ratio for that context of use. However, they should then be aware of the accompanying privacy concerns and address these by informing the user about what kind of biometric data is stored or how it is protected. Furthermore, an alternative solution for users that refuse to provide their biometric data should be sought.

\section{Related Work}

This section first describes related work on how user perceptions differ between authentication schemes. Second, we explore the factors influencing user perceptions in general, and how the context of use influences the users' perceptions of authentication schemes in particular. 

\subsection{User Perceptions of Authentication Schemes}

 As described above, authentication schemes are usually divided into knowledge-based, token-based, and biometric schemes \cite{grassi2017digital}. While knowledge-based (e.g., a password) and token-based schemes (e.g., a chip card) have been around for a while, also biometrics (e.g., fingerprint recognition) are increasingly available to end-users on a large scale through technological advances, such as cameras and fingerprint sensors in smartphones \cite{bhagavatula2015biometric}. This has given rise to comparisons of biometrics with knowledge-based and token-based schemes: For example, Furnell and Evangelatos \cite{furnell2007public} found that 61\% of survey participants preferred biometrics to knowledge-based and token-based schemes. Yet, there were also differences between different types of biometrics, e.g., fingerprint, iris, and retina authentication were rated as more reliable than voice, keystroke, or signature authentication. Contrary to that, users felt uncomfortable using iris or retina authentication as compared to fingerprint, hand, and signature authentication. 
 
 The results of a focus group by Dörflinger \textit{et al.} \cite{dorflinger2010my} revealed similar ambiguities: Users perceived retina and fingerprint authentication as most secure for use on smartphones. However, the intention to use fingerprint authentication was highest, and that to use retina scans lowest among all tested schemes.
 Instead of biometrics, in another focus group and online study by Ben-Asher \textit{et al.} \cite{ben2011need} more participants intended to use knowledge-based PIN/password authentication as compared to biometrics on their smartphones. Yet, fingerprint authentication was perceived as more secure compared to PIN/password authentication. 
 
 A laboratory study 
 \cite{zimmermann2019password} studying authentication on a laptop revealed similar conflicting perceptions in terms of password and fingerprint authentication. While password authentication was preferred for use, fingerprint authentication was rated as most secure. Furthermore, a nearly equal number of participants expressed their highest and lowest preference for fingerprint authentication in a ranking. 
 
 A study by Ruoti et al. \cite{ruoti2015authentication} compared the user perceptions of seven authentication schemes in two rounds. They first compared email-based single sign-on, federated single sign-on (using passwords as authentication schemes), and token-based schemes within their authentication categories. Then the three ``winners'' were compared between authentication categories. The usability was rated highest for the password-based Google OAuth 2.0 and the token-based approach Snap2Pass, in which a QR-code is scanned with a paired smartphone. 

While our approach is similar to that of Ruoti et al. \cite{ruoti2015authentication} in terms of selecting and comparing the ``winners'' of previous authentication studies, it differs in that our research includes the category of biometric schemes. Furthermore, while Ruoti et al. \cite{ruoti2015authentication} designed tasks in which users signed in to a fictional forum versus bank account to include different information sensitivity levels, they did not compare the user perceptions of the schemes across these contexts explicitly. Our study bridges that gap by explicitly studying user perceptions of all three schemes across four contexts of use.

 \subsection{Influences on User Perceptions}
 
Among the factors influencing acceptance of authentication schemes researchers have identified perceived ease of use (e.g., \cite{ho2003biometric}), perceived usefulness (e.g., \cite{al2009effects}), perceived invasiveness of the scheme (e.g., \cite{ho2003biometric}), social influences (e.g., \cite{de2015feel}), and perceived privacy and security (e.g., \cite{morosan2012voluntary}). 
 For example, the privacy concerns mentioned as problematic with regards to biometrics \cite{coventry2003honest,de2015feel,riley2009culture} might be a reason for the polarization expressed in the above-mentioned laboratory study \cite{zimmermann2019password} and in the study by Riley \textit{et al.} \cite{riley2010security}.
 
In line with the identification of perceived ease of use as a potential predictor for acceptance \cite{ho2003biometric}, other studies suggest a strong correlation between perceived usability and preference  \cite{riley2010security,zimmermann2019password}. Several studies \cite{riley2010security,riley2009culture,zimmermann2019password} also discussed habit or familiarity, respectively, as a potential factor influencing user perception. 
 Finally, some studies also found indications for the influence of personal and demographic variables, such as security expertise \cite{wolf2019pretty}, age \cite{qiu2019towards}, gender \cite{sieger2012gender}, and culture \cite{riley2009culture}. 
 
 
 

 Overall, these findings indicate that perceptions and preferences should be considered when choosing or designing authentication schemes. This research, therefore, comprehensively compares the influence of relevant perceptions as suggested by previous work (e.g., perceived usability, security, and privacy) as well as the users' familiarity with different schemes on the users' preference and intention to use the respective schemes.





\subsection{Influence of the Authentication Context}

Furnell and Evangelatos \cite{furnell2007public} also explored the users' attitudes towards using biometrics for different purposes and found that users felt more comfortable with using biometrics for passports as compared, e.g., to staff ID cards. Heckle \textit{et al.} \cite{heckle2007perception} found that participants were more inclined to use fingerprint authentication to protect personal information for private purchases as compared to corporate purchases.
However, an interview study by Riley \textit{et al.} \cite{riley2010security} comparing perceptions of using biometrics at different locations including an airport, rail station, and retail environment did not reveal clear differences. 

Abbott and Patil \cite{abbott2020mandatory} found mandatory use of two-factor authentication in a university did not decrease user acceptance and experience for sensitive applications, but for other, less sensitive accounts.

In a survey comparing user preferences for password and biometric authentication across different contexts of use, such as online banking and social networks, users universally expressed higher preferences for passwords \cite{gerber2017security}. Yet, the study also revealed significant differences between the users' general preference and three specific contexts of use: social networks, cloud services, and email accounts. 
Likewise, Jones \textit{et al.} \cite{jones2007towards} found that the percentage of respondents finding passwords, tokens, and biometrics useful and acceptable varies with the context of use.

These findings suggest that the context of use well plays an important role, especially in terms of the type of service and type of data accessed \cite{abbott2020mandatory,furnell2007public,gerber2017security,heckle2007perception,jones2007towards}. This research thus compares user perceptions across different types of services that grant access to different types of data.

\section{Method}
This section first describes the selection of the three authentication schemes followed by the study procedure, a description of the sample, and a short discussion of ethical considerations.

\subsection{Selection of Authentication Schemes}
\begin{figure}[ht!]
  \centering
  \includegraphics[width=.8\columnwidth]{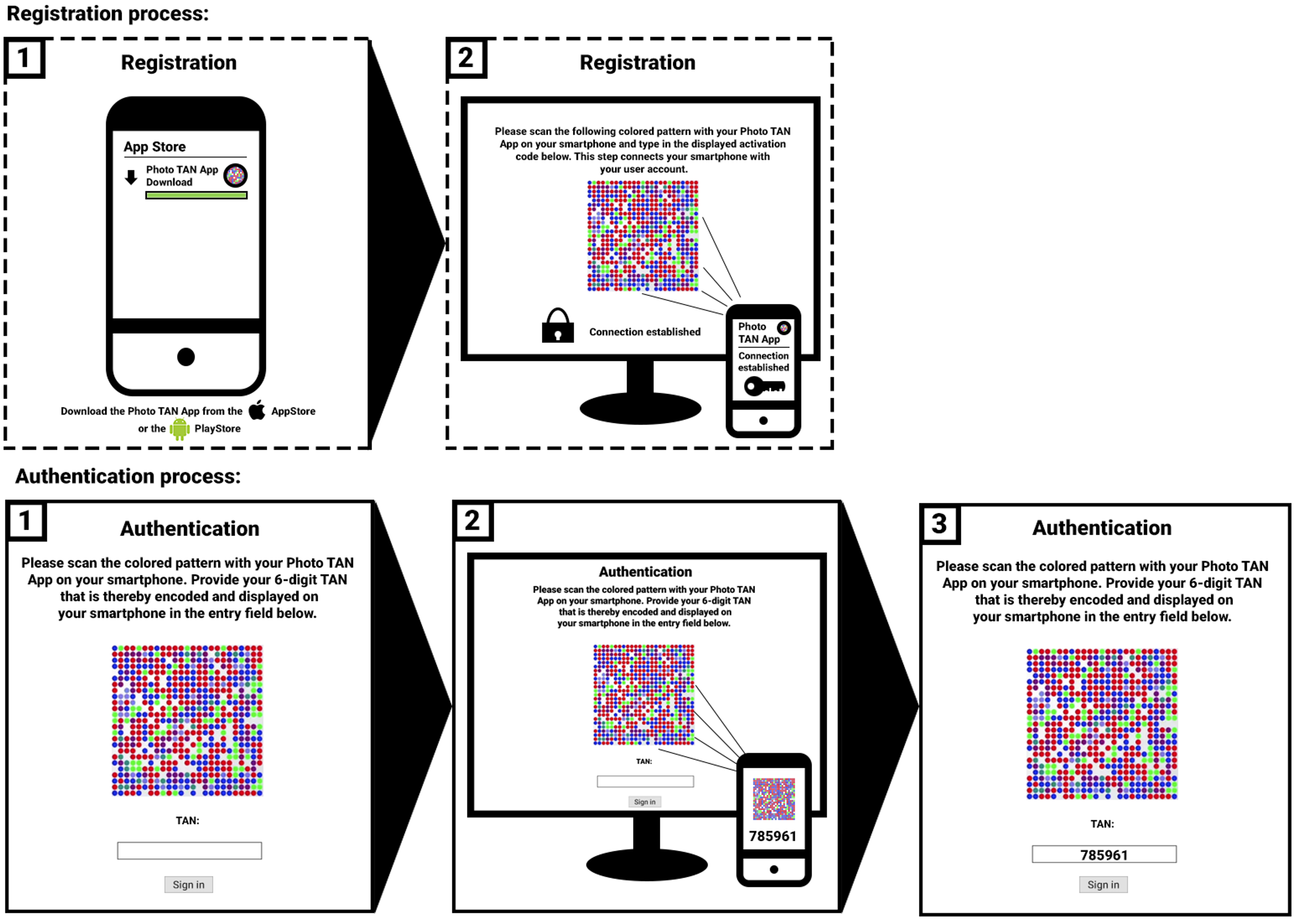}
  \caption{Graphical depiction of the registration and authentication process of the smartphone-based scheme as used in the study.}~\label{fig:figurePhotoTAN}
\end{figure}

This research compared perceptions of the password, fingerprint, and a smartphone-based scheme. 
For password authentication, the user memorizes and types in an alphanumeric ``secret'' that is only shared between the system and the user.
Fingerprint authentication is a biometric scheme. To authenticate certain points of a user's fingerprint that are stored in the system are compared with the fingerprint captured by a sensor upon login. 
The smartphone-based scheme developed by OneSpan \cite{onespan19} is a token-based scheme in which a smartphone application generates an individual, secret key that is stored on the smartphone. Only with the unique key, and thus the smartphone, a numeric code can be extracted from a coloured pattern dynamically created for each authentication process. The numeric code that changes with every authentication trial can be used to log into an account or to grant a financial transaction. The scheme is often used in the banking sector labelled as PhotoTAN scheme. The registration and authentication process of the smartphone-based scheme is graphically depicted in Figure \ref{fig:figurePhotoTAN}.

The three schemes were chosen for the following reasons:

\subsubsection{Representatives of Relevant Authentication Categories} 
The most common classification of authentication schemes is into knowledge-based (something you know), biometric (something you are), and token-based schemes (something you have). As it is unfeasible either in terms of the number of required participants, cost, or length of the study to evaluate a large number of authentication schemes against each other and each within different contexts of use, we decided to include a representative of each authentication category. Similar to the procedure suggested by Ruoti et al. \cite{ruoti2015authentication}, we chose the three ``winning'' schemes based on previous research.
With the password, fingerprint, and smartphone-based scheme a representative of each category has been included to account for differences in the user perceptions related to the features of the authentication scheme and the secret it is based on.

\subsubsection{Previous Research on User Perceptions}
In a previous laboratory study \cite{zimmermann2019password} \textit{N}=41 participants interacted with mock-ups of a total of 12 
different authentication schemes that had been selected based on a rating of authentication schemes. 
Evaluating the interaction with these schemes in terms of several aspects, the password, fingerprint, and smartphone-based scheme scored high across the measured perceptions, especially in terms of preference, perceived usability, security readiness, and intention to use. Similarly, they scored low in terms of perceived effort and problem expectation. We chose to include the three ``winners'' within their authentication category because of their relevance and to minimize differences based on preference beforehand (even though these cannot be excluded completely), and instead to maximise differences between contexts of use and authentication categories as such. 

 \subsubsection{Practical Relevance} 
 Besides theoretical and methodological considerations, the practical relevance of the study for researchers, developers and providers was important for us. With the password, fingerprint, and smartphone-based scheme, we included schemes that are currently frequently used. The password still is the most commonly applied authentication scheme \cite{authentication18} and a relevant fallback mechanism. The fingerprint is the most frequently used biometric \cite{authentication18}. Integrated fingerprint sensors in laptops and smartphones facilitate access for a large number of users.
 The smartphone-based scheme is a spreading token-based scheme, especially in the banking sector. With the new European Payment Services Directive (PSD2) \cite{eu-2366} and the clause for secure customer authentication, many banks have moved from SMS-based one-time codes to other authenticator applications such as the smartphone-based scheme analysed in this study\cite{cimpanu19}. It furthermore bears similarities to other smartphone-based schemes using QR-codes such as Snap2Pass \cite{dodson2010secure}. 
 The scheme does not require providers or users to distribute costly tokens but allows using the already carried smartphone for authentication. 
 
 Overall, the high scores in terms of user preference and intention to use in previous research also demonstrate the practical relevance of the schemes from a user perspective. Furthermore, as representatives of different authentication categories, the schemes also offer potential for combining two of these into two-factor authentication schemes to further increase authentication security.  

\subsection{Procedure}

\begin{figure*}[h!]
  \centering
  \includegraphics[width=1\columnwidth]{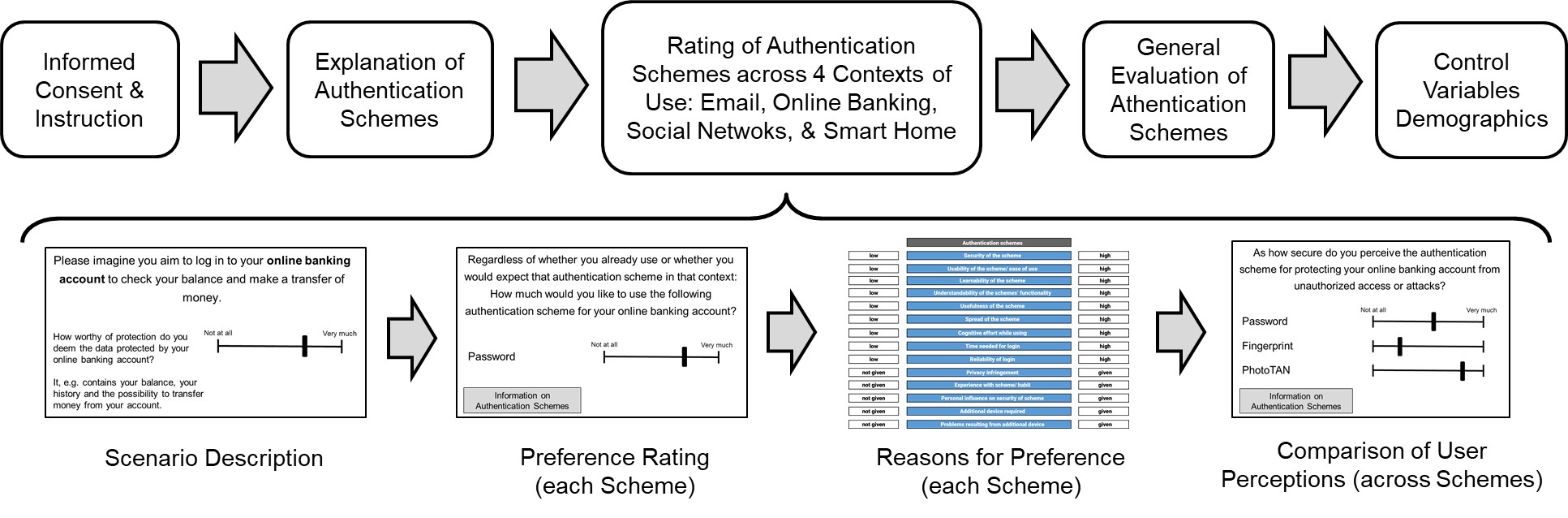}
  \caption{Graphical depiction of the study procedure.}~\label{fig:figureprocedure}
\end{figure*}

To provide an overview, the procedure is depicted in Figure \ref{fig:figureprocedure}. 
On the first page of the online survey, the participants were presented with an informed consent sheet. If they decided to proceed, they received an explanation of all three authentication schemes that were analyzed in the study. The descriptions consisted of a short text description accompanied by a series of images that detailed the registration and login process for each scheme. As an example, the visual description of the smartphone-based scheme used in the study is shown in Figure \ref{fig:figurePhotoTAN}. The participants were then asked to rate their familiarity with a scheme as this was found to potentially impact on their evaluation in previous research \cite{weir2009usable,zimmermann2019password} and the frequency with which they used it on a visual analogue scale ranging from 1 to 100. The schemes were presented in a randomised order to control for sequential effects. 

The main part of the study consisted of the rating of the three authentication schemes in different contexts of use. These included authentication for an email account, social networks, online banking, and smart homes. The selection was based on choosing relevant accounts used by a large number of people that protected different types of data, such as personal and work communication (email), social data (social networks), financial data (online banking) and usage data related to the own home and devices (smart home). The context smart home, even though not used by the majority of people yet, was included due to its increasing spread \cite{smarthome19} and importance. The number of analyzed contexts was limited to four after a pilot test to not overly increase the duration of the online study, which would potentially decrease the quality of the data. At the same time, the within-subject design was chosen to be able to compare the data between the contexts for the same group of people.

For each context of use, we controlled whether people actually used that context, if yes, which services they used, and how sensitive they deemed the data that was protected in that context. 
People were then provided with a short scenario description and asked to imagine they were in that situation, e.g., ``Imagine you'd like to log into your online banking account to check your balance and to make a transfer of money.''

Following the scenario description, the participants were asked to rate how much they liked to use the password, the fingerprint scheme or the smartphone-based scheme for authentication on a 100-point scale, and to provide reasons for their choice. For all contexts of use, the participants rated all three authentication schemes. 
A list of potential reasons for the participants' choice was derived from a literature review described in \cite{zimmermann2019keep} 
and previous interview studies including the follow-up interviews in \cite{zimmermann2019password}. 
The potential reasons were clustered in terms of whether they concerned the login credentials or the authentication scheme and displayed as a matrix. Figure \ref{fig:mcshort} shows an exemplary excerpt; the complete matrix is included in the Supplementary Material. By using this matrix, we aimed to facilitate participants' responses and to be able to explore the applicability and importance of the different aspects quantitatively. Thus, while our previous interviews followed an inductive approach to derive potential reasons, this study used a deductive approach based on the findings. 
The participants were asked to check all aspects that had been relevant for their decision, and to indicate whether their choice was due to the aspect being given or not given, or being high or low respectively. Participants had the possibility to describe further aspects not included in the matrix.

The participants were then asked to rate the perceived security, privacy, effort-benefit ratio of all three authentication schemes, and their intention to use a certain scheme within a context on a visual analogue scale ranging from 1 to 100. The type of scale was chosen to ensure the collection of metric data \cite{reips2008interval,sung2018visual}.

The final part of the study consisted of a general comparison of all three schemes, and questionnaires asking for technological affinity, general privacy concerns and socio-demographic information. The three schemes were compared in terms of perceived security, privacy concerns, perceived effort-benefit ratio, and efficiency on continuous scales ranging from 1 to 100. Subjective usability was compared using the System Usability Scale (SUS) \cite{brooke1996sus} as also suggested by Ruoti et al. \cite{ruoti2015authentication} as an appropriate metric for the usability of authentication schemes. We used the complete scale except for item 5 (integration of different functions). To avoid distraction and possibly distorted scores due to a lack of applicability of this item for some authentication schemes such as the password it was decided to leave it out. Lewis and Sauro \cite{lewis2017can} successfully showed that any one item of the SUS can be left out without having a significant effect on the resulting scores. To account for the missing item, the overall scores have been adapted according to Lewis and Sauro's suggestion \cite{lewis2017can}.

Further, trust in the authentication schemes was measured applying the ``Facets of System Trustworthiness'' (FOST) scale by Franke \textit{et al.} \cite{franke2015advancing} in a version adapted for hypothesized use. An additional item for error-proneness was included. 
Technological affinity was measured with the ``Affinity for Technology Interaction'' (ATI) scale \cite{franke2019personal}, privacy concerns were measured using a scale for global information privacy concerns from Smith \textit{et al.} \cite{smith1996information} and adapted by Malhotra \textit{et al.} \cite{malhotra2004internet}.

\subsection{Sample}
A total of \textit{N}=201 German participants took part in the study. Of these, \textit{N}=98 identified as female, \textit{N}=103 as male, and \textit{N}=1 as other/diverse. All participants were aged over 18 with the following distribution: \textit{N}=25 were between 18 and 24, \textit{N}=86 were between 25 and 34, \textit{N}=22 were between 45 and 54, and \textit{N}=15 were between 55 and 64 years old. In terms of education, \textit{N}=87 had finished school, \textit{N}=30 had completed an apprenticeship or other training, and \textit{N}=85 held a university degree. Some participants (\textit{N}=19) reported their occupation being related to IT (Security). 

The technological affinity of the sample was \textit{M}=3.91 (\textit{SD}=.95) out of 6 points. The sample's general privacy concerns were \textit{M}=4.80 (\textit{SD}=.82) out of 7 points.

The participants were recruited via the Online Survey Panel Clickworker \cite{clickworker} with random sampling within 
the country's population to achieve heterogeneity in terms of the participants' gender, age, education and location. The study took about 30 minutes to complete, and participation was compensated according to the country's minimum wage. 

\begin{figure}[ht!]
  \centering
  \includegraphics[width=0.8\columnwidth]{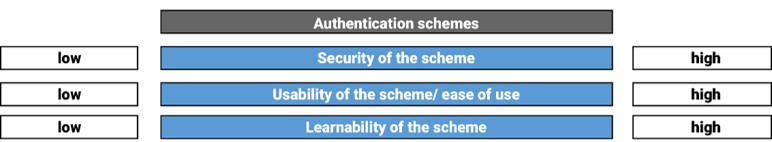}
  \caption{Exemplary extract of the qualitative rating matrix.}~\label{fig:mcshort}
\end{figure}

\subsection{Ethical Considerations}
The study was planned and conducted in line with the ethics checklist provided by our university's ethics committee and guidelines for ethical psychological research \cite{Ethics05}. All participants were presented an informed consent sheet containing information about the aim of the study, the procedure and the expected duration. Further, participants received information about the handling of their data in accordance with EU data protection laws, their rights in terms of data privacy and the contact information of the researchers. The survey was implemented in SoSci Survey \cite{leiner2014sosci} that stores data in the EU in accordance with strict EU data protection laws. Further, the collection of demographic information was reduced to a minimum of relevant control variables: gender, age, education and occupation. Age was further collected in age ranges to increase anonymity. The participants were compensated in line with the survey tool's suggestions and based on the country's minimum wage.   

\section{Results}
This section first describes the preparation of data followed by a detailed analysis of all four hypotheses and the qualitative analysis of the decision factors. It closes with an overview on the familiarity with the schemes and exploratory analyses.

\begin{table*}[ht!]
  \centering
  \caption{Descriptive values of the user perceptions of different authentication schemes in different contexts of use. Note: The scale ranged from 0 to 100, GE = General, EM = Email, SN = Social Network, OB = Online Banking, SH = Smart Home, M = Mean, SD = Standard Deviation.
  \label{tab:Tabelle1}}{
  \includegraphics[width=1\columnwidth]{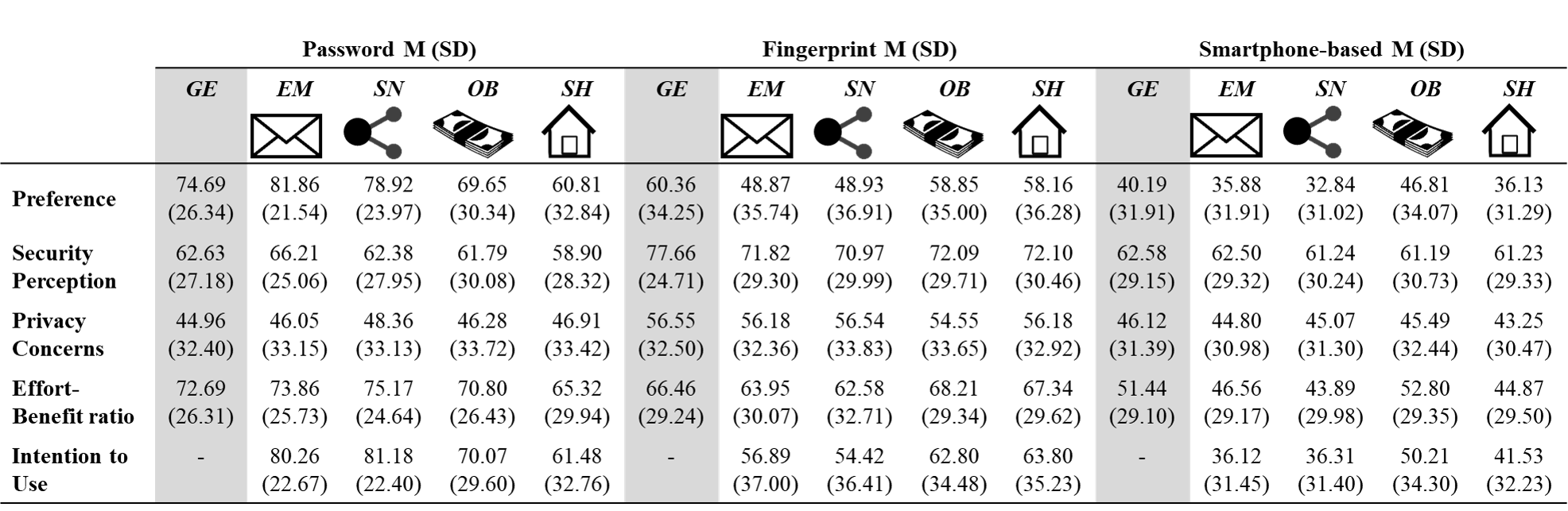}}
\end{table*}

\subsection{Data Preparation and Analysis}
Before the analysis, all incomplete (\textit{N}=27) and implausible responses (\textit{N}=3) were removed, leading to \textit{N} = 201 remaining participants for the analysis. As implausible responses, we classified the cases that provided the same rating, e.g., 100 on a 100-point scale, across all questions indicating that the instructions and items had not been read. 

The analysis was conducted with IBM SPSS Version 24 and R Version 4.0.1 \cite{R}. All tests were conducted on a significance level of $\alpha$ = .05. To address existing deviations from normality, robust procedures were used whenever available following the explanations of Field and Wilcox \cite{field2017robust} as well as Mair and Wilcox \cite{mair2019robust}. These are further detailed in the related sections. Furthermore, where appropriate, multiple testing was accounted for using the Bonferroni procedure.

\subsection{H1: Differences in User Perceptions across Schemes}

Table \ref{tab:Tabelle1} shows the descriptive values of the general rating in terms of preference, perceived security, privacy concerns, and effort-benefit ratio (continuous scales from 1 to 100). Furthermore, the general perceived efficiency (continuous scale from 1 to 100), the trust values (measured with the FOST scale plus an additional item on 6-point scales), and usability ratings (measured with the SUS on a scale from 0 to 100), are shown in Table \ref{tab:general}. The SUS scores have been adjusted to compensate for the missing items as suggested by Lewis and Sauro \cite{lewis2017can}.

For analysing differences in general user perceptions of authentication schemes within-subjects, first a repeated-measures, multivariate ANOVA (RM MANOVA) was conducted. Using Pillai's trace\footnote{Pillai's trace or the Pillai-Bartlett trace respectively has been chosen as a measure because it has been found to be most robust to violations of assumptions such as normality when sample sizes are equal \cite{bray1985multivariate} as cited by \cite{field2013discovering}}, there was a significant effect of the authentication schemes on the users' perceptions, \textit{V}=.65, \textit{F}(14,182)=24.00, \textit{p}<.001.



\begin{table}
  \centering
  \caption{Descriptive values of the general ratings of the three authentication schemes. Note: M = Mean, SD = Standard Deviation. The efficiency and usability scale ranged from 1 to 100, the trust scale from 1 to 6.\label{tab:general}}{%
  \begin{tabular}{l r r r}
    &  \multicolumn{3}{c}{\small{\textbf{Authentication Schemes M(SD)}}} \\
    \cmidrule(r){2-4}
    {\small\textit{Measure}}
    & {\small \textit{Password}}
      & {\small \textit{Fingerprint}}
    & {\small \textit{Smartphone-based}} \\
    \midrule
   \small Efficiency & \ 78.55 (22.72) & 79.75 (23.22) &  62.92 (28.05) \\
    \small Usability & 79.73 (16.41) & 73.76 (20.16) & 54.59 (20.11) \\
    \small Trust & 4.28 (.52) & 4.25 (.61) & 4.08 (.67) \\
    \hline
  \end{tabular}}
 \end{table}

To follow up that finding univariate analyses were conducted. Even though the results of the RM MANOVA post-hoc tests and that of robust repeated-measures ANOVAs were very similar, the results of the robust analysis are reported considering deviations from normality in some variables. The robust repeated-measures ANOVAs were conducted using the WSR2 package in R \cite{mair2019robust}. The tests were specified with the default trimming rate of .2 and post-hoc test \textit{p}-values were compared to Bonferroni-corrected significance levels. 
 
 The results revealed that the users' general perceptions of authentication schemes differed significantly for all measured perceptions, i.e. the dependent variables general preference, security perception, privacy concerns, effort-benefit ratio, efficiency, usability, and trust. The results of the analyses, including the results of the pairwise comparisons, are depicted in Table \ref{tab:H1robust}. Significant results after comparing the results with the Bonferroni-corrected \textit{p}-values are marked in bold. 
 
 The table shows that all comparisons of the fingerprint and smartphone-based scheme were significant. Furthermore, security perception, as well as privacy concerns, were significantly higher for fingerprint authentication as compared to the password and the smartphone-based scheme. Usability and preference ratings were highest for the password, followed by the fingerprint and smartphone-based scheme.

\begin{table*}
  \centering
   \caption{Comparison of the general ratings of the three authentication schemes using robust RM ANOVA. Note: PW = Password, FP = Fingerprint, SB = Smartphone-based, EB ratio= Effort-benefit ratio  F = Test statistic, df = degrees of freedom,  p = p-value (significant p-value applying Bonferroni correction marked in bold), partial $\omega^{2}$ = effect size.\label{tab:H1robust}}{%
  \begin{tabular}{l r r r r r r r r r r r}
    &  \multicolumn{5}{c}{\small{\textbf{Robust Repeated Measures ANOVA}}} &  \multicolumn{2}{c}{\small{\textbf{PW - FP}}} &  \multicolumn{2}{c}{\small{\textbf{FP - SB}}} &  \multicolumn{2}{c}{\small{\textbf{SB - PW}}}\\
    \cmidrule(r){2-6}  \cmidrule(r){7-8} \cmidrule(r){9-10} \cmidrule(r){11-12}
    {\small\textit{Measure}}    & {\small \textit{F}} & {\small \textit{{df c}}}    & {\small \textit{df e}} & {\small \textit{p}}  & {\small \textit{par. $\omega^{2}$}}  & {\small \textit{$\hat \psi$}} & {\small \textit{p}}  & {\small \textit{$\hat \psi$}}  & {\small \textit{p}}  & {\small \textit{$\hat \psi$}}  & {\small \textit{p}}\\
    \midrule
   \small Preference & \ 38.37 & 1.98 & 205.61 &  \textbf{<.001} & .26 & 11.181 & \textbf{.005}   & 20.476 & \textbf{<.001}  & 34.295 & \textbf{<.001}  \\
   \small Security Perception& \ 24.97 &  1.91 & 199.03 & \textbf{<.001} & .18 & -19.181 & \textbf{<.001} & 14.524 & \textbf{<.001} & -3.800 &  .283 \\
   \small Privacy Concerns & \ 8.25  &  2.00 & 207.7 & \textbf{<.001} & .06 & -8.371 & \textbf{.005}  & 7.867 &  \textbf{.004} & -0.762 & .786  \\
    \small Effort-Benefit Ratio & \ 24.37 &  2 & 208 & \textbf{<.001} & .18 & 4.752 & .119   & 14.361 & \textbf{<.001} & 23.371  &  \textbf{<.001}  \\
    \small Efficiency & \ 28.41 &  1.92 & 199.38 & \textbf{<.001} & .21  & -2.105 & .292 & 13.648 &  \textbf{<.001} & 13.895 & \textbf{<.001} \\
    \small SUS Score & \ 106.55 &  1.98 & 205.7 & \textbf{<.001} & .50 & 4.497 & \textbf{.001}  & 17.910 & \textbf{<.001} & 22.804 & \textbf{<.001}  \\
    \small Trust Score & \ 6.53 & 1.78 & 185.52 & \textbf{.002} &  .05 & -0.021 & .745  & 0.142 & \textbf{<.001}  & 0.167 & \textbf{.012} \\
    \hline
  \end{tabular}}
 \end{table*}


\subsection{H2/H3: Differences in User Perceptions across Context of Use}

Table \ref{tab:Tabelle1} shows the descriptive values for each context of use in terms of preference, security perceptions, privacy concerns, effort-benefit ratio, and intention to use. The values include both the perceptions of people stating to use the context and people not actually using the context but asked to imagine using the context. Further results for each context of use are described in the following:

\subsubsection{Email Account}
Of the 201 participants, 185 (91.58\%) used an email account. The most frequently used services were GMX (\textit{N}=60), Web (\textit{N}=49), Outlook (\textit{N}=48), Googlemail (\textit{N}=32), and Thunderbird (\textit{N}=23). A total of 34 participants reported using more than one email account. Email account sensitivity was rated as \textit{M}=76.73 (\textit{SD}=23.17) on a 100-point scale.

\subsubsection{Online Banking}
A total of \textit{N}=175 participants had at least one online banking account. Online banking data was rated as very sensitive with \textit{M}=88.07 (\textit{SD}=21.50).

\subsubsection{Social Network}
Social Networks were used by 151 participants. The most often used social networks were Facebook (\textit{N}=116), Instagram (\textit{N}=65), Twitter (\textit{N}=20), and Xing (\textit{N}=13). The protect-worthiness of the data  was rated with \textit{M}=70.50 (\textit{SD}=28.29).

\subsubsection{Smart Home}
 A total of 33 participants stated to live in a smart home (SH) or to own smart home devices. The most commonly named technologies or devices were Amazon Alexa/Echo (\textit{N}=21), Google Home (\textit{N}=4) and Homeatic (\textit{N}=2). The sensitivity of SH data was rated with \textit{M}=73.12 (\textit{SD}=31.66). Because of the small number of people already using smart home devices, we tested whether the sensitivity rating differed between people owning and people not owning smart home devices. A Mann-Whitney-U test revealed no significant differences between smart home owners (\textit{M}=77.91, \textit{SD}=23.83, \textit{Md}=84.00) and non-smart home owners (\textit{M}=73.04, \textit{SD}=32.31, \textit{Md}=88.00) with \textit{Z}=-0.057, \textit{p}=.955. 

\subsubsection{H2: Perceptions of schemes across contexts of use}

To evaluate the perceptions of the individual schemes across the four contexts of use, repeated measures MANOVAs were conducted. Using Pillai's trace, there was a significant effect of the context of use on the users' perceptions of the password scheme, \textit{V}=.37, \textit{F}(15,180)=7.07, \textit{p}<.001, partial $\eta^{2}$=.37. Similar results were found for the fingerprint scheme, \textit{V} = .20, \textit{F}(15,180) = 2.95, \textit{p} <.001, partial $\eta^{2}$ = .20, and the smartphone-based scheme, \textit{V} = .32, \textit{F}(15,175) = 5.36, \textit{p} <.001, partial $\eta^{2}$ = .32. 

Similar to the reasoning for analysing $H_1$, robust repeated-measures ANOVAs were conducted to look for univariate differences in the dependent measures individually. Again, the test was done using the WSR2 package in R \cite{mair2019robust} with the default trimming rate of .2. The results are presented in Table \ref{tab:H2robust}. The table shows that the perceived effort-benefit ratio, preference for and the intention to use a certain scheme differed significantly across the four contexts of use for each of the schemes. However, the users' perceptions of security and privacy of a scheme did not vary across contexts of use.

\begin{table*}
  \centering
  \small
  \caption{Differences in perceptions of individual schemes across contexts of use analysed with robust repeated-measures ANOVAs. Note: EB ration= Effort-benefit ration, F = Test statistic, df = degrees of freedom, p = p-value (significant p-value applying Bonferroni correction marked in bold), partial $\omega^{2}$ = effect size.\label{tab:H2robust}}{%
  \begin{tabular}{l r r r r r r r r r r r r r r r}
    &  \multicolumn{5}{c}{\small{\textbf{Password}}} &  \multicolumn{5}{c}{\small{\textbf{Fingerprint}}} &  \multicolumn{5}{c}{\small{\textbf{Smartphone-based}}} \\
    \cmidrule(r){2-6}  \cmidrule(r){7-11}  \cmidrule(r){12-16} 
      & {\small \textit{F}} & {\small \textit{df c}}    & {\small \textit{df e}} & {\small \textit{p}}  & {\small \textit{$\omega^{2}$}}   & {\small \textit{F}} & {\small \textit{df c}}    & {\small \textit{df e}} & {\small \textit{p}}  & {\small \textit{$\omega^{2}$}}    & {\small \textit{F}} & {\small \textit{df c}}    & {\small \textit{df e}} & {\small \textit{p}}  & {\small \textit{$\omega^{2}$}}\\
    \midrule
   \small Preference & \ 23.44 & 2.29 & 238.53 &  \textbf{<.001} & .18 & 14.17 & 2.79 & 289.87 &  \textbf{<.001} & .11 & 12.73  & 2.62 & 272.33 &  \textbf{<.001} & .10 \\
   \small Security & \ 2.25 &  2.84 & 295.32 & .086 & .01 & 1.56 & 2.92 & 303.17 &  .200 & .01  & 0.04 & 2.86 & 297.78 &  .989 & .00  \\
   \small Privacy & \ 0.92  &  2.95 & 306.47 & .432 & .00  & 0.64 & 2.99 & 310.99 &  .589 & .00  & 0.63 & 2.94 & 305.88 &  .595 & .00  \\
    \small EB Ratio & \ 6.17 &  2.70 & 280.90 & \textbf{<.001} & .05 & 4.45 & 2.95 & 306.97 &  \textbf{.005} & .03 & 7.85 & 2.89 & 300.86 &  \textbf{<.001} & .06 \\
    \small Intention & \ 28.78 &  2.24 & 232.58 & \textbf{<.001} & .21 & 8.84 & 2.95 & 306.77 &  \textbf{<.001} & .07  & 18.48 & 2.74 & 284.91 &  \textbf{<.001} & .14\\
    \hline
  \end{tabular}}
     \end{table*}

\subsubsection{H3: Interaction of schemes and contexts of use}

At the time of the analysis and to our awareness, there was no non-parametric or robust alternative for analysing the interaction effect of the two independent within-subject factors context and scheme on one outcome variable (two-way repeated measures ANOVA) or numerous outcome variables (two-way repeated-measures MANOVA) \cite{field2012discovering}. 
This section thus reports the results of a factorial repeated measures MANOVA including the scheme and the context of use as independent variables. Despite the use of Pillai's trace as a relatively robust measure for deviations from normality in equally-sized samples, the central limit theorem, and similar results provided by the classical and robust measures in the previous analyses, the results should be interpreted cautiously given the deviations from normality in some of the variables. For a discussion of this limitation, potential reasons for the deviations, and ideas for future work, please see the limitations section. 

The factorial repeated measures MANOVA revealed that the interaction effect of the context of use and the authentication scheme on the users' perceptions was significant, Pillai's trace \textit{V} = .46, \textit{F}(30, 159) = 4.44, \textit{p} <.001, partial $\eta^{2}$ = .46. Also, the main effects of the context and the scheme on the user perceptions were significant,  Pillai's trace \textit{V} = .24, \textit{F}(15, 174) = 3.62, \textit{p} <.001, partial $\eta^{2}$ = .24, and Pillai's trace \textit{V} = .65, \textit{F}(10, 179) = 32.94, \textit{p} <.001, partial $\eta^{2}$ = .65 respectively.

Follow-up analyses revealed that context influenced the participants' perceptions of the effort-benefit ratio (\textit{F}(2.78, 522.60) = 4.39, \textit{p} =.006,  partial $\eta^{2}$ = .02), intention to use (\textit{F}(2.83, 531.96) = 7.70, \textit{p} <.001,  partial $\eta^{2}$ = .04) and preference (\textit{F}(2.73,512.83) = 11.50, \textit{p} <.001,  partial $\eta^{2}$ = .06), but not privacy (\textit{F}(2.89, 542.62) = .41, \textit{p} =.738,  partial $\eta^{2}$ = .002) or security perceptions (\textit{F}(2.91,547.79) = 1.82, \textit{p} =.145,  partial $\eta^{2}$ = .01). 
The type of scheme, however, significantly influenced all measured perceptions, including security and privacy. The test results are shown in Table \ref{tab:H3_1}.

\begin{table}
  \centering
    \caption{Differences in user perceptions of schemes across contexts of use. Note: EB ration= Effort-benefit ratio, F = Test statistic, df = Greenhouse-Geisser corrected degrees of freedom, p = SPSS Bonferroni corrected significance value,  $\eta^{2}$ = partial $\eta^{2}$ effect size.\label{tab:H3_1}}{
  \begin{tabular}{l r r r r r}

    {\small\textit{Measure}}
    & {\small \textit{F}}  & {\small \textit{df c}}    & {\small \textit{df e}} & {\small \textit{p}}  & {\small \textit{ $\eta^{2}$}}     \\
    \midrule
   \small Preference &  79.27  & 1.97 & 371.06 & <.001 & .30\\
    \small Security & 11.36 & 1.96 & 367.75 & <.001 & .06 \\
    \small Privacy & 12.77 & 1.96 & 368.53 & <.001 & .06 \\
       \small EB ratio &  61.50  & 1.98 & 372.43 & <.001 & .25\\
          \small Intention &  60.90  & 1.98 & 372.30 & <.001 & .25\\
    \hline
  \end{tabular}}
  \end{table}

\begin{figure*}[h!]
  \centering
  \includegraphics[width=0.9\columnwidth]{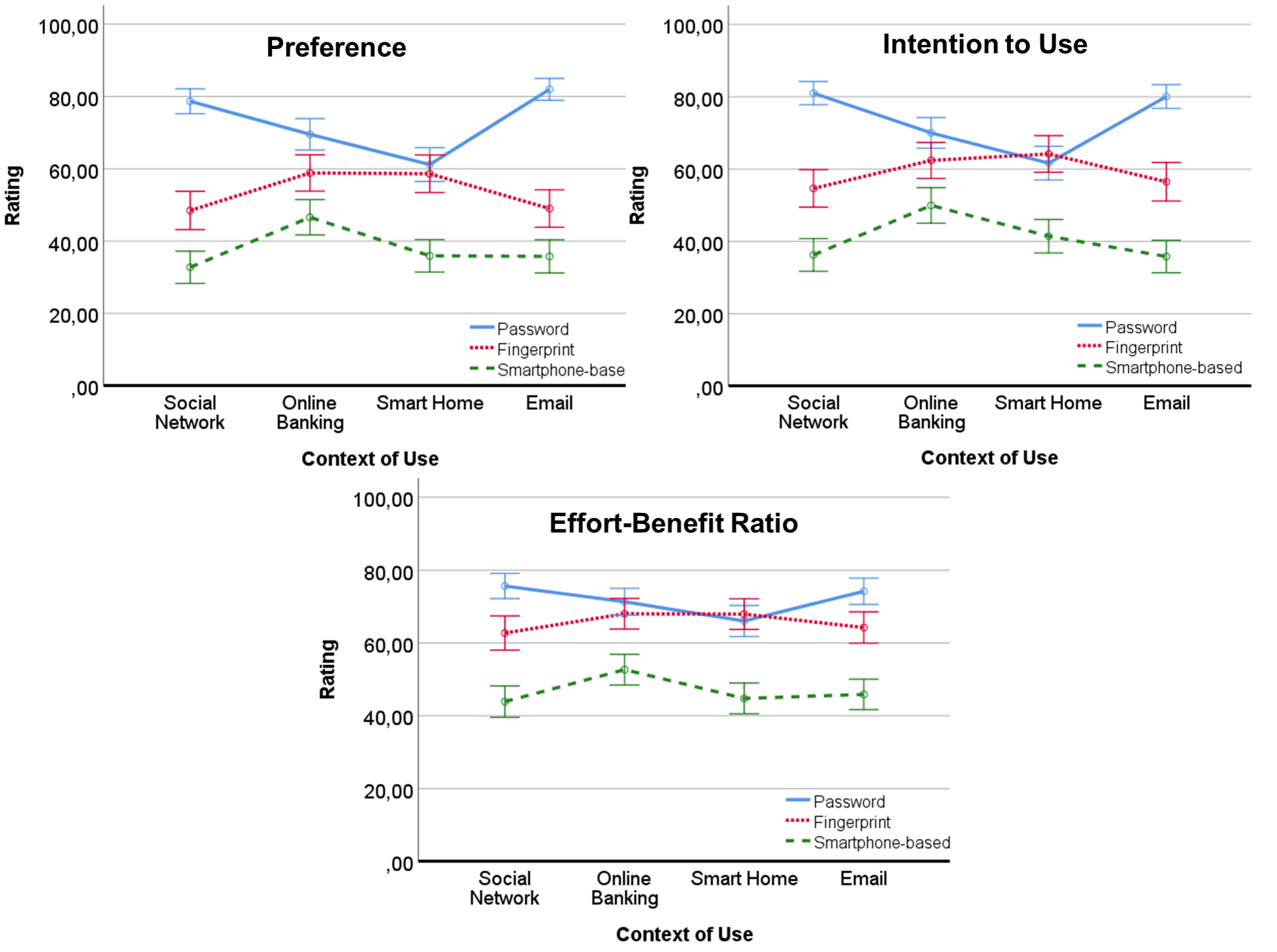}
  \caption{Graphical depiction of the interaction effects in terms of the perceived effort-benefit ration, intention to use, and preference}~\label{fig:graph}
\end{figure*}

Finally, the interaction of the context of use and the authentication schemes influenced the perceived effort-benefit ratio (\textit{F}(5.60, 1052.79) = 8.15, \textit{p} <.001,  partial $\eta^{2}$ = .04), intention to use (\textit{F}(4.98, 935.39) = 30.40, \textit{p} <.001,  partial $\eta^{2}$ = .14), and preference (\textit{F}(4.97, 935.10) = .12, \textit{p} <.001,  partial $\eta^{2}$ = .12), but not the security (\textit{F}(5.41, 1016.98) = 1.61, \textit{p} =.15,  partial $\eta^{2}$ = .01) and privacy perception (\textit{F}(5.50, 1034.04) = .48, \textit{p} =.48,  partial $\eta^{2}$ = .01). Figure \ref{fig:graph} shows the interaction effects in terms of the effort-benefit ratio, the intention to use a certain scheme and the preference for a certain scheme. 

\subsection{H4: Predictors for Authentication Preference}

To analyse to what extent different aspects influence the preference for a certain scheme within a certain context of use, we conducted a quantitative analysis as well as a rating of qualitative features.

\subsubsection{Quantitative analysis}
Throughout the study, the participants rated each scheme in general and within each context of use in terms of different aspects on a 100-point scale. Using these measures, multiple regressions with preference as a dependent variable were conducted. As predictors, we included the following variables in the model: 
\begin{itemize}
    \item (1) variables measures for the scheme and context: the perceived security, perceived privacy concerns and perceived effort-benefit ratio for using a certain scheme within a certain context
    \item (2) variables measured for the context: the protect-worthiness of the data in a specific context
    \item (3) variables measured for the scheme: perceived efficiency, SUS score, trust score (measured on a 6-point scale), familiarity with, and frequency of use of a certain scheme
\end{itemize}

Overall, twelve multiple regressions were calculated across the three schemes and four contexts of use. All predictors were entered into the models simultaneously. After an initial scan of the data, cases with residuals larger than 3.29 (outside the 99.9\% limit of a normal distribution) have been deleted. This concerned one case each for the models preference for fingerprint authentication in the context smart home and preference for password and the smartphone-based authentication in the context social networks.  

Furthermore, to account for deviations from normality, robust confidence intervals and significance tests have been calculated using a bootstrapping function with 1000 samples and 95\% bias-corrected confidence intervals. Additionally, robust regression coefficients have been calculated using a robust regression via the ``Essentials for R'' plug-in in SPSS. 
The initial and robust model coefficients, the robust confidence intervals and significance tests of all multiple regressions are presented in the Supplementary Material. 
In the following, we summarize the models, the main findings of the analysis, and differences between schemes and contexts of use as the focus of this analysis. 

Table \ref{tab:H4_1} displays model summaries of the twelve multiple regressions. It shows that all models predict the preference for using a certain scheme in a certain context of use significantly. The models, i.e. the predictors in the models, account for 30 to 57\% of the variation in preference.

\begin{table}
  \centering
    \caption{Summary of the multiple regressions. Note: $R^2$ = Measure of variance accounted for by the model, F = Test statistic, p = significance value.\label{tab:H4_1}}{
  \begin{tabular}{l r r r }
    {\small\textit{Dependent Variable}}
    & {\small \textit{$R^2$}}
  & {\small \textit{F}}    & {\small \textit{p}}   \\
    \midrule
\multicolumn{4}{c}{\small{\textbf{Email Account}}} \\
\midrule
\small Preference for Password	&	0.39	&	12.47	&	<.001	\\
\small Preference for Fingerprint	&	0.43	&	14.68	&	<.001	\\
\small Preference for Smartphone-based scheme	&	0.34	&	9.99	&	<.001	\\
   \midrule
\multicolumn{4}{c}{\small{\textbf{Social Network}}} \\
\midrule							
\small Preference for Password	&	0.34	&	8.98	&	<.001	\\
\small Preference for Fingerprint	&	0.35	&	10.77	&	<.001	\\
\small Preference for Smartphone-based	&	0.40	&	12.66	&	<.001	\\
   \midrule
\multicolumn{4}{c}{\small{\textbf{Online Banking}}} \\
\midrule						
\small Preference for Password	&	0.57	&	25.38	&	<.001	\\
\small Preference for Fingerprint	&	0.41	&	13.51	&	<.001	\\
\small Preference for Smartphone-based	&	0.47	&	17.05	&	<.001	\\
   \midrule
\multicolumn{4}{c}{\small{\textbf{Smart Home}}} \\
\midrule						
\small Preference for Password	&	0.38	&	12.16	&	<.001	\\
\small Preference for Fingerprint	&	0.53	&	22.33	&	<.001	\\
\small Preference for Smartphone-based	&	0.30	&	8.29	&	<.001	\\
    \hline
  \end{tabular}}
     \end{table}

On the level of the individual predictor variables, Table \ref{tab:H4_2} provides an overview of the significant regression coefficients calculated using multiple regression and a robust multiple regression approach to account for deviations from normality. The robust estimates are given in brackets. The remaining regression coefficients and additional data such as standard errors and confidence intervals for each estimate can be found in the Supplementary Material. 

From Table \ref{tab:H4_2}, it becomes visible that perceived security and effort-benefit ratio for using a scheme within a certain context are relevant predictors across different schemes and contexts of use. On the level of the scheme, the SUS score impacts preference values across multiple schemes and context. Contrary to that, the privacy concerns for using a certain scheme within a certain context, the general perceived efficiency of the scheme, the frequency of use and familiarity with a certain scheme are only significant predictors for one or two models respectively. Furthermore, the standard regression coefficients are not substantially different from the robust estimates, yet in some cases, only the standard or only the robust estimate was significant. 

\begin{table*}
  \centering
    \caption{Overview of the significant predictors in the multiple regressions. Note: The trust score values are different from the other coefficients as they are based on 6-point instead of a 100-point scale, \textit{p} < .05,  b = regression coefficient, $b_r$ = robust regression coefficient.\label{tab:H4_2}}{%
  \begin{tabular}{l r r r r r r r r r r r r}
& \multicolumn{3}{c}{\small{\textbf{Online Banking   b($b_r$)}}} & \multicolumn{3}{c}{\small{\textbf{Social Network b($b_r$)}}} &\multicolumn{3}{c}{\small{\textbf{Smart Home b($b_r$)}}} &\multicolumn{3}{c}{\small{\textbf{Email Account b($b_r$)}}}\\
    \cmidrule(r){2-4}  \cmidrule(r){5-7}  \cmidrule(r){8-10} \cmidrule(r){11-13} 
& \textit{PW}	&	\textit{FP}	&	\textit{SB}	&	\textit{PW}	&	\textit{FP}	&	\textit{SB}	&	\textit{PW}	&	\textit{FP}	&	\textit{SB}	&	\textit{PW}	&	\textit{FP}	&	\textit{SB}	\\
    \midrule
\small Security &	.56	&	.40 	&	.39 	&	-	&	-	&	-	&	.43 	&	.45	&	.20	&	-	&	.24	& -		\\
 &	(.65)	&	(.52)	&	(.49)	&	-	&	(.21)	&	(.19)	&	 (.50)	& (.54)	&	 (.19)	&-		&	 (.25)	&	-	\\
\small Privacy Concerns &	(-.11)	&	-	&	-	&	-	&-		&	-	&	-	&	-	&	-	&	-	&	-	&-		\\
\small Effort-Benefit Ratio  &	.33 	&	.23 	&	.24 	&	.32 	&	.41 	&	.50 	&	.35 	&	.39 	&	.34	&	-	&	.47	&	.53 	\\
 &	(.32)	&	(.23)	& (.24)	&	 (.24)	&	 (.50)	&	(.56)	&	 (.37)	&	 (.41)	&	(.42)	&	-	&	(.50)	&	 (.62)	\\
\small Protect-worthiness &	-	&	.43 	&	-.26	&-		&	-	&	-	&	.22 	&	-	&	-	&	.22 	&	-	&	-	\\
 &	-	& (.39)	&-	&	-	&	-	&	-	&	 (.20)	&	-	&	-	&	(.23)	&	-	&	-	\\
\small Efficiency & -	&	-	&	-	&	(.17)	&	-	&	-	&	-	&-		&	-	&	(.23)	&	-	&-	\\
\small SUS Score &-	&	.37 	&	.34 	&	.22 	&	.32 	&	-	&	-	&	.29 	&	.40 	&	.25 	&	.31	&	-	\\
 &-	&	(.46)	&	(.34)	&	(.23)	&	(.23)	&		&	-	&	(.23)	&	 (.42)	&	.(.31)	&	-	&	-	\\
\small Trust Score	&-	&-	&	-	&	-	&	-	&	-	&	-	&	10.42 	&	-	&-		&	10.35 	&	-	\\
	&-	&	(8.96)	&	-	&	-	&	-	&	-	&	-	&	(10.50)	&	-	&	-	&	 (10.25)	&	-	\\
\small Frequency of use		&-	&	-	&	-	&	(.26)	&	-	&	-	&	-	&	-	&	-	&	-	&	-	&-		\\
\small Familiarity	&-	&	-	&	-	&	(-.30)	&-		&	-	&	-	&	-	&	-	&	-	&	-	&	-	\\
    \hline
  \end{tabular}}
    \end{table*}

\subsubsection{Rating of Qualitative Features}

\begin{table*}[h!]
  \caption{Frequencies of preference groups for every authentication scheme and context. Note: EM = Email, SN = Social Network, OB = Online Banking, SH = Smart Home.\label{tab:MC_freq}}{%
  \centering
  \includegraphics[width=1\columnwidth]{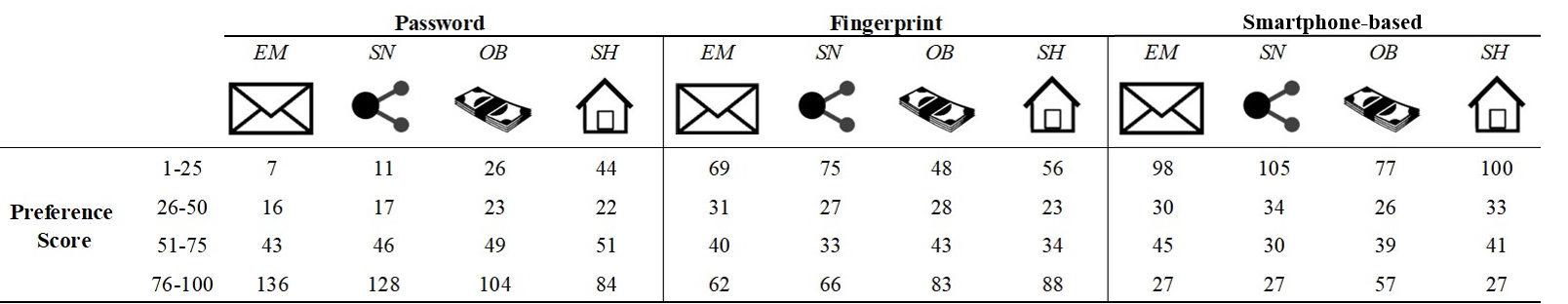}}
\end{table*}



First of all, this section provides a brief overview of the methodological approach of the qualitative analysis. The following sections then summarize typical reasons for or against aiming to use a specific authentication scheme for a certain context of use and also deal with context-specific peculiarities. 

In the course of the study, all participants had the opportunity to choose reasons for their preference for all three authentication schemes for each of the four contexts of use (email account, online banking, social networks and smart home). The selection was carried out using a dichotomous matrix with a total of 18 different properties (see Figure \ref{fig:mcshort} for an exemplary excerpt). 

For the analysis of the selected reasons, we considered the participants who had either a very high ($\leq$ 75) or a very low ($\geq$ 25) preference score on a 100-point scale for the respective authentication scheme in a certain context. Table \ref{tab:MC_freq} summarizes the frequencies of participants per authentication scheme in each context and preference group. For the derivation of the main reasons for or against the respective authentication scheme, only those reasons were considered that were given by at least 80\% of the respective subgroup. A breakdown of all 18 reasons for each scheme and context can be found in the Supplementary Material. Figure \ref{fig:mcauswahl} provides an exemplary excerpt of the data used for this analysis. It shows that the participants explained their preference for using each scheme for the email account varies due to personal experience, the perceived personal influence on the scheme and expected problems with an additional device (if required).

\subsubsection{Password}
For participants who indicated a high preference for the password scheme, a high level of perceived security, the scheme's spread, perceived usability, little time needed for login, and personal influence on the choice of login credentials were the main reasons, although the cognitive effort for remembering the login credentials, i.e. the password, is acknowledged. 

Particularly in the context of online banking, the personal influence on the security level of the scheme and the fact that no additional device is required for authentication are additional reasons. 

Typical reasons for the rejection of the password scheme included the perceived high effort for remembering the login credentials and a perceived low security level, although the scheme was nevertheless perceived as cognitively less complex, requiring little time for login, easy to use, widespread, and easy to learn. However, the small sample of rejecting participants (\textit{N} < 30), especially in the email and social networks context, should be noted. 


\subsubsection{Fingerprint}

For participants who indicated a high preference for the fingerprint scheme, the focus is on good usability, ease of learning, little time required for login, and little cognitive effort. In addition, the procedure is perceived as secure. 

Typical reason for the rejection of the fingerprint scheme include the perceived low spread of the scheme, and thus a lack of personal experience with the authentication scheme. Furthermore, a lack of personal influence on the login credentials, and the high privacy concerns associated with the provision of the personal fingerprint, as well as the necessity of an additional device and possible problems resulting from that were rated as problematic.  Nevertheless, the procedure was found to be very usable with low cognitive and temporal effort and perceived as secure, even by users rejecting the scheme.

\subsubsection{Smartphone-based scheme}

For participants who indicated a high preference for the smartphone-based scheme, the main reasons were the ease of learning, good usability, low cognitive effort, high reliability and the high level of security of the procedure. It should be noted that in this group considerably more participants (in relation to those with negative preferences) stated that they already had experiences with the procedure.
However, this group also acknowledged the necessity and possible problems with the additional device, i.e. the smartphone, required. 

Typical reasons for the rejection of the smartphone-based procedure included a high expenditure of time, poor usability, as well as the need for and problems with the additional device required. Particularly in the online banking context, the perceived lack of the scheme's spread and, consequently, the lack of experience with the procedure were selected as additional factors. 

\begin{figure*}[h!]
  \centering
  \includegraphics[width=1\columnwidth]{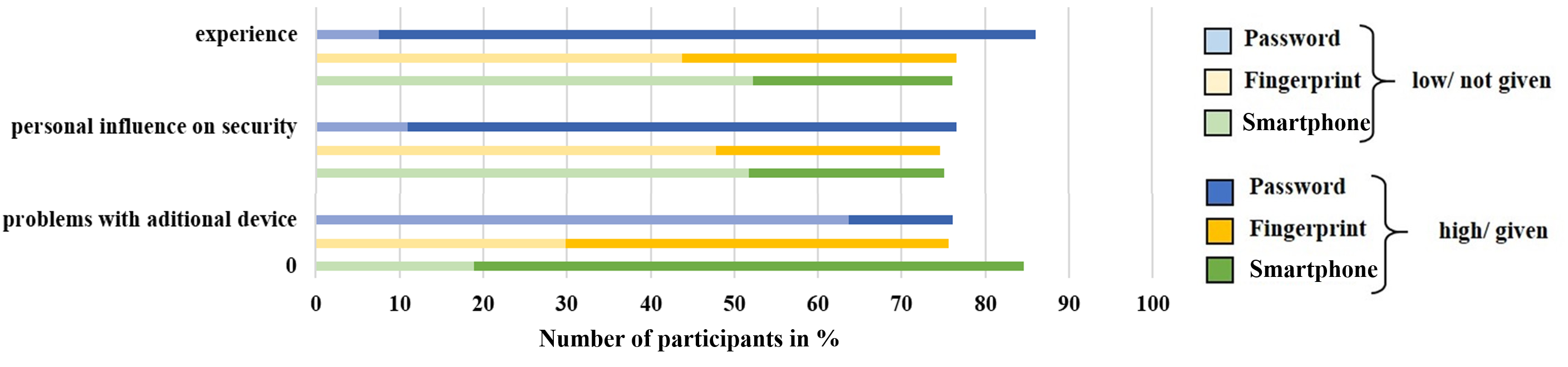}
  \caption{Graphical depiction of exemplary reasons for the participants' preference ratings in terms of the email account}~\label{fig:mcauswahl}
\end{figure*}

\subsection{Familiarity with the Schemes}
The familiarity with the password scheme was rated highest  with \textit{M}=93.18 (\textit{SD}=14.95) out of 100 followed by the fingerprint (\textit{M}=66.36, \textit{SD}=30.55) and the smartphone-based scheme (\textit{M}=42.45, \textit{SD}=36.09). The frequency of use followed a similar pattern: Password (\textit{M}=85.01, \textit{SD}=19.37), Fingerprint (\textit{M}=26.87, \textit{SD}=29.71), Smartphone-based scheme (\textit{M}=20.82, \textit{SD}=28.09).

\subsection{Exploratory Analyses}

Asked for other, not analysed, schemes for the different contexts of use, the participants most frequently mentioned voice recognition for the smart home context (\textit{N}=8) followed by iris (\textit{N}=6) and face recognition (\textit{N}=4). The need for MFA, especially in combination with some kind of one-time password or code was most often expressed in the context of online banking (\textit{N}=24), also followed by iris (\textit{N}=5) and face recognition (\textit{N}=4). The most frequently mentioned alternative to the analysed schemes in terms of social networks was face recognition (\textit{N}=8), whereas in terms of the email account participants again most frequently mentioned one-time passwords or codes (\textit{N}=6).

\section{Discussion}

This section will first summarize the findings in comparison with the results of related work. It will then discuss implications resulting from the findings and summarize relevant recommendations for decision-makers. The section ends with a discussion of limitations and suggestions for future work. 

\subsection{Summary of Findings}

The results show that type of authentication scheme and the context of use individually, and in combination, impact the users' perceptions of different authentication schemes confirming $H_1$, $H_2$, and $H_3$. 

Overall, password authentication, despite its downsides in terms of cognitive load, still scored highest in terms of the users' preference and perceived usability. The SUS score assigned to passwords in our study (\textit{M}=79.37, \textit{SD}=16.41) is similar to that assigned to the password-based federated single sign-on service evaluated by Ruoti et al. \cite{ruoti2015authentication} in their final evaluation round (\textit{M}=75.00, \textit{SD}=14.8). That fingerprint authentication scored significantly higher in terms of perceived security but was also accompanied by the highest level of privacy concerns, confirms the ambiguity in terms of the user perceptions already shown in previous research \cite{cherapau2015impact,de2015feel,riley2010security,zimmermann2019password}. This was also evident in the qualitative responses provided by our participants. Another explanation is offered by the familiarity with the schemes: Password authentication that most participants were familiar with and used frequently received the highest preference ratings, whereas the smartphone-based scheme that is relatively new in comparison and that people were much less familiar with, received the lowest preference ratings.

Still, the finding is not in line with previous findings  by 
Ruoti et al. \cite{ruoti2015authentication}. Ruoti et al. \cite{ruoti2015authentication} found a high preference for the smartphone-based scheme and a scheme labelled Snap2Pass \cite{dodson2010secure} that bears similarities to the smartphone-based scheme analysed in this study, respectively. In Ruoti et al.'s study, even though many participants were not familiar with the Snap2Pass scheme either, it was perceived as innovative and attributed a certain ``coolness'' factor. Participants in the final evaluation round assigned a mean SUS score of \textit{M}=68.4 (\textit{SD}=16.7) to Snap2Pass while participants in our study rated the smartphone-based scheme significantly lower with \textit{M}=54.59 (\textit{SD}=20.11). An explanation for this difference might lie in the study design: In Ruoti et al's \cite{ruoti2015authentication} 
and Zimmermann and Gerber's \cite{zimmermann2019password} studies participants actually tested the scheme and perceived the ``coolness'' factor first-hand rather than seeing a visual description. This explanation is supported by the fact that people that qualitatively expressed a high preference in this study more often had experience with the scheme than people that expressed a negative attitude. 
Another explanation might be the demographics of the sample. While this study comprised a heterogeneous sample with different age groups, \cite{ruoti2015authentication} and
\cite{zimmermann2019password}  mainly studied undergraduate students that might be more open for using new technologies.  

Apart from that, our study revealed significant differences across different contexts of use, types of services and data, respectively. It is noticeable that the preference for and the intention to use a certain scheme as well as the effort-benefit ratio differed significantly across contexts of use, while the security and privacy perception of the schemes did not. The security and privacy perceptions thus seem to be linked to the scheme and to remain relatively stable across different contexts. This indicates that users somehow trade-off the perceived security level with the context of use and perhaps other factors. It also indicates that maximum perceived security and privacy protection are not the only reasons for users' preference ratings. 

Another indication for the impact of familiarity on user  perceptions can be seen in the differences across contexts of use: The password preference and intention to use ratings were highest for social networks and email accounts, contexts in which password authentication has been typically used for a long time. Likewise, smartphone-based authentication scored highest in terms of online banking, the context it is currently frequently deployed in. The results in terms of the smart home context, a context that few people have actual experience with, are less clear, e.g., fingerprint and password authentication received nearly equal ratings. 

However, when modelling the impact of the familiarity with the scheme and frequency of use on the preference ratings in multiple regressions, the two variables did not appear as significant predictors (except for one case). Several explanations might account for these ambiguous findings in terms of familiarity and frequency of use: For example, familiarity and frequency of use were measured on a general level only once, but not individually for each context of use. It might thus be that context-specific differences have been lost in the analysis. Furthermore, the familiarity and frequency of use data were constrained in that, e.g., the majority of people was highly familiar with the password scheme but there were only a few data points at the lower end of the scale. This density at the upper end of the scale with a mean value above 90 out of 100 might have made it difficult to calculate a regression coefficient. Future research looking into the impact of familiarity on user perceptions should therefore consider using context-specific measures that are unconstrained in that the scale is more evenly used by the participants. 

Apart from that, the findings in terms of familiarity and frequency of use have implications for the final hypothesis. $H_4$ can only be partially confirmed as some but not all of the expected variables could be shown to be significant predictors for the preference to use a certain scheme within a certain context. While the perceived security and effort-benefit ratio, as well as the scheme's general SUS score, seem to be relevant, positive predictors across different schemes and contexts, other predictors seem to be more specific. For example, the protect-worthiness of the data affects the preference for password authentication in the smart home and email account context. 
Furthermore, the trust scores were only relevant for predicting the preference for fingerprint authentication in the smart home and email account context. 

Finally, while all models significantly predicted preference for using a certain scheme within a certain context explaining between about 30 and 60\% of the variation in preference, no clear pattern of differences between contexts or schemes emerged. The remaining 40 to 70\% of unexplained variance indicates that there are other variables not measured in this research that influence user preference ratings. These might provide a clearer picture of context- and scheme-specific differences. Potential candidates include measures that lie in the person such as security expertise or personality, or additional context-specific information such as the frequency with which people use certain types of accounts. 

\subsection{Implications}

Overall, the results reveal that, like with the often researched technical and security aspects, there is no ``silver bullet'' in terms of user perceptions. However, the findings provide valuable insights into how user perceptions differ across three popular authentication schemes and four contexts of use. In addition, the study identified some factors that do influence user perceptions across schemes and contexts of use, and some factors that do less so. The results have several implications for research and practice that will be discussed in more detail in the following. The findings can be used to inform the following aspects:

\begin{itemize}
   \item the selection of authentication schemes
   \item the implementation and design of selected authentication schemes 
   \item the design of new context-dependent or context-aware authentication schemes and mechanisms
\end{itemize}

\subsubsection{Selection of authentication schemes}

When selecting an authentication scheme, organisations or service providers often face certain requirements. This can include compatibility with existing hardware or software, cost, or requirements in terms of a certain security level. Aiming to support researchers and practitioners in identifying authentication schemes meeting the requirements amongst a plethora of schemes, choice support systems such as ACCESS \cite{mayer2016supporting,mayer2018accessv2,renaud2014access} have been developed. Yet, the support system does not provide the ``best'' scheme, but calculates a performance value for the schemes meeting the requirements. To choose among a range of suitable schemes, the results of this research can provide insights into which type of scheme is deemed acceptable for certain types of accounts or data from a user perspective. 

For schemes or contexts outside this investigation, or for identifying suitable combinations of schemes for MFA, the findings provide support which measures to investigate to make an informed decision. From this research, it seems that perceived security, usability, and effort-benefit ratio are especially relevant. 

Finally, the findings may also encourage decision-makers to provide a selection of schemes to the users. For example, the fingerprint scheme at the same time is rated as highly secure and invoking privacy concerns. Taking into account this ambivalence decision-makers aiming to implement fingerprint authentication might consider providing a second alternative for people concerned for their privacy. 

\subsubsection{Implementation of Authentication Schemes}

Regardless of whether an authentication scheme has been selected considering user perceptions or not, e.g., because using a certain scheme is mandatory, the findings can still inform how an authentication scheme is rolled out and how the user interface is designed. For example, the password scheme and the smartphone-based scheme have been rated lower in terms of the security perception than the fingerprint scheme. This perception could be addressed, e.g., by informing users about the security features of the smartphone-based scheme before its implementation, by making use of adequate ``security theatre'' \cite{schneier2008psychology}, or by providing users with support for creating secure passwords such as a password meter in the interface (e.g., \cite{carnavalet2015large,ur2017design}). 
Furthermore, perhaps existing misconceptions in terms of the schemes can be detected beforehand and addressed in the design of the interface. For example, while fingerprint authentication is perceived as very secure, it may be a good idea to make users aware that its security, if available, also depends on the security of the fallback mechanism. As many smartphone applications, for example, rely on a PIN as a fallback mechanism, users should be informed and encouraged to also select a secure password or PIN.  

\subsubsection{Design of Context-dependent Authentication}

The findings of this study can also be used to inform the design of or extend context-aware authentication concepts such as the ones presented by Hayashi \textit{et al.} \cite{hayashi2013casa}, Hulsebosch \textit{et al.} \cite{hulsebosch2005context} or the risk-based approach to strengthen password authentication evaluated by Wiefling et al. \cite{wiefling2020evaluation,wiefling2020more}. For example, the user ratings in terms of security or effort-benefit ratio, or related concepts such as ``information security readiness'' \cite{sun2008users,sun2011more}, can be used to evaluate the effort people are willing to take for differing levels of data sensitivity. Instead of all-or-nothing authentication, the security level might be increased from no authentication for non-sensitive information to a password for more sensitive ones and the combination with an additional fingerprint as MFA for very sensitive applications. This approach may be applicable for a variety of contexts such as smartphone applications \cite{dorflinger2010my,hayashi2012goldilocks} ranging from the display of the weather forecast to online banking applications, or smart home functionalities \cite{ashibani2017context} ranging from turning on the light to health data or camera access.

\subsection{Recommendations for Decision-makers}
From the discussion above, the following implications for decision-makers can be summarized:

\begin{itemize}
    \item Even after applying organisational, security-related or legal requirements, decision-makers might still be met with a ``tie'' between different authentication schemes to choose from. Decision-makers should consider users' perceptions (especially perceived security, usability, and effort-benefit ratio) within a given context to break that tie in favor of their user group and to increase acceptance.
    \item Decision-makers deciding to implement biometric schemes should consider the privacy concerns held by a relevant user group. They should either address the concerns by the type of implementation or provide an alternative scheme to not exclude or upset a relevant group of users.
    \item Beyond the selection of the scheme, user perceptions should be considered in the system design. Potential misconceptions in terms of privacy or security might negatively affect the usage behavior and actual authentication security. Thus, user perceptions, e.g. in terms of the security of fall-back mechanisms or in terms of what makes a secure password, should be addressed and secure behavior should be supported. 
    \item The results in terms of the perceived sensitivity of data in different contexts can be used to inform the design of context-dependent authentication to optimize the perceived effort-benefit ratio for the user, and perhaps even for the service-provider offering the authentication mechanism.   
\end{itemize}

\subsection{Limitations}
\label{limitations}

It is likely that the familiarity with the smartphone-based scheme has increased since the study due to the European Payment Services Directive 2 (PSD2) \cite{eu-2366} that came into effect in September 2019 and among other things requires strong authentication and the dynamic generation of new TANs for each transaction. This led some major banks in the country where the study took place to offer the smartphone-based scheme as a replacement for previous authentication schemes. Therefore, it would be interesting to follow up on that development and to see how the increased use of the smartphone-based scheme in the banking sector, also known as PhotoTAN scheme in that context, affects the users' perceptions. 

In the data analysis, we faced deviations from normality that are plausible content-wise, but even with a large sample size and equally sized groups might impact tests assuming (multivariate) normality such as ANOVA or MANOVA. Examples include skewness in terms of the protect-worthiness of online-banking data or in terms of the preference for password authentication. To mitigate the effects, we chose robust test statistics such as Pillai's trace, strict Bonferroni correction for multiple testing, and applied robust alternative procedures where possible. Even though the results of the robust and classical approach were similar, a bias in the analysis of the third hypothesis where no robust procedure was available cannot be completely excluded. 
Alternative options such as transforming the data (e.g., logarithmize the data) were considered and tested, yet did not yield the effect hoped for. Furthermore, the need to apply different transformations to different variables within the same model would have caused serious difficulties in interpreting the differences.

Finally, the distribution of the preference ratings especially caught our attention and provided relevant questions for future research: While the password was well and the smartphone-based scheme less preferred, i.e., both were skewed, no clear preference in terms of the fingerprint scheme emerged. While there seems to be a group of people highly preferring the fingerprint scheme, others reject it. This very well confirms the split in terms of the biometric procedure already visible in previous research. From our quantitative and qualitative data, privacy concerns and previous experience with the scheme offer a potential explanation for the finding. Future research should consider including these and other potentially relevant grouping variables and separately analyse people with a low versus high preference for the fingerprint scheme or perhaps other biometrics as well.  

\subsection{Conclusion}
To shine light on the seemingly intractable issue of authentication choice, this research systematically compared user perceptions of three different authentication schemes, representing different authentication categories, across different contexts of use. In a within-subject online study \textit{N}=201 participants rated each scheme in four contexts of use in terms of preference, usability, security perceptions, privacy concerns, effort-benefit ratio, and intention to use. In addition, users selected the reasons for their preference ratings in a matrix including potential reasons based on prior research. 

Despite the users acknowledging the cognitive load, password authentication received the highest preference ratings across all contexts of use. Fingerprint authentication was relatively most preferred in the online-banking and smart home context. Yet, the users' perceptions of the fingerprint schemes were ambivalent: It was perceived as very secure, but at the same time raised privacy concerns. This tension likely led to a split in the users' fingerprint preferences ratings. The smartphone-based scheme was mostly preferred in the online banking sector, a context it currently increasingly deployed in following a new EU directive \cite{eu-2366}. 

The results further revealed that the type of scheme and the context of use individually, and in combination, impact user perceptions of a certain scheme. That security and privacy perceptions differed across schemes, but not contexts, indicates that these perceptions are closely tied to the scheme. It is thus likely that security and privacy are somehow traded-off with other aspects such as usability when it comes to the preference of using a scheme in a certain context of use. 

The findings have implications not only for the selection and implementation of authentication schemes, but also provide relevant insights for designing user-centred and context-aware authentication schemes.

\section{Acknowledgement}

This research work has been funded by the German Federal Ministery of Education and Research and the Hessen State Ministry for Higher Education, Research and the Arts within their joint support of the National Research Center for Applied Cybersecurity ATHENE, and by the Deutsche Forschungsgemeinschaft (DFG, German Research Foundation) – 251805230/GRK 2050.




\bibliographystyle{acm}
\bibliography{sample}

\end{document}


\maketitle

\section{Table of Contents}

\begin{itemize}
    \item Description of the registration and log-in procedure of the authentication schemes
    \item Scenario descriptions
    \item List of items - Context of Use
    \item List of items - Overall evaluation
    \item Control variables and demographics
    \item H4: Reasons for authentication preference sorted by context of use 
    \item H4: Differences in reasons for authentication preference 
    \item H4: Summary of the Multiple Regressions 
\end{itemize}

\section{Description of the registration and log-in procedure of the authentication schemes}

\subsection{Information text on Authentication}

With the help of an authentication scheme users can prove a purported feature or identity respectively. Based on the result access to certain information can be granted or denied. One example is password authentication. e.g. for an email account. By logging in with an email address and an associated password the user proves his or her access authorization for this email account. The authentication scheme thus protects personal data from unauthorized access by third parties.

In the following we will present three authentication schemes. Please read the information carefully.
Afterwards we will ask you to rate the presented authentication schemes.

\subsection{Password authentication}

Password authentication is a \textit{knowledge-based} authentication scheme. Upon registration the user creates a secret word a a sequence of alphanumeric symbols  (e.g. letters. numbers. special characters). The secret word or sequence needs to be memorized and typed in upon log-in.

The following figure \ref{fig:password} visualizes the registration (left) and authentication procedure (right). The registration process needs to be completed once. Afterwards. only the  authentication process takes place upon every log-in. 

\begin{figure*}
  \centering
  \includegraphics[width=1\columnwidth]{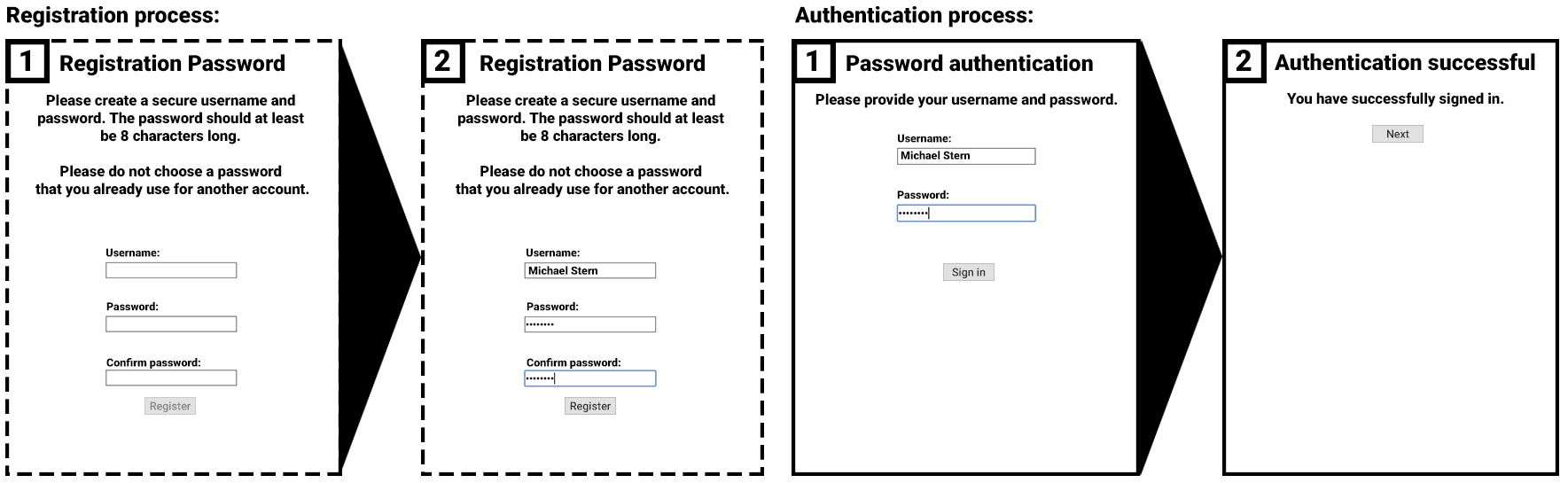}
  \caption{Graphical depiction of the registration an authentication process of password authentication.}~\label{fig:password}
\end{figure*}

\subsection{Fingerprint authentication}

The authentication via fingerprint is a \textit{biometric} authentication scheme. The user is authenticated via a unique, physical feature, the fingerprint. Upon registration the users' fingerprint or certain parts or points thereof are captured via a sensor. Upon log-in the stored information is compared to the information of the fingerprint.

The following figure \ref{fig:fingerprint} visualizes the registration (left) and authentication procedure (right). The registration process, in this case the capturing and storing of the fingerprint information, needs to be completed once. Afterwards. only the  authentication process takes place upon every log-in.

\begin{figure*}
  \centering
  \includegraphics[width=1\columnwidth]{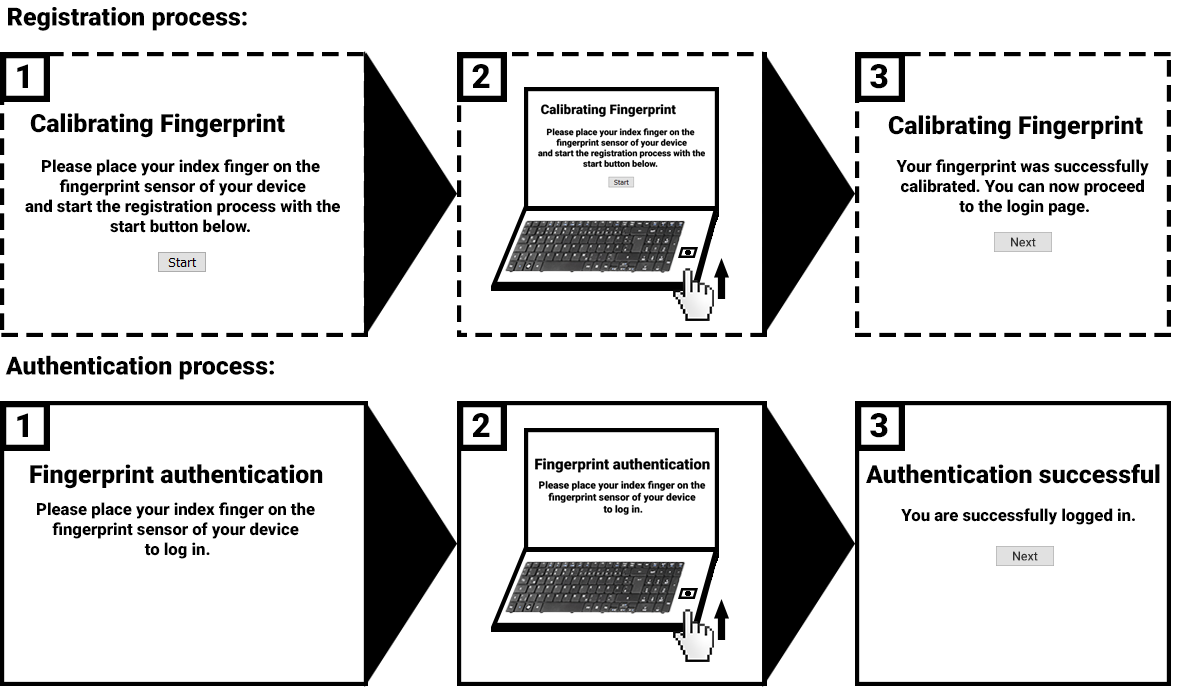}
  \caption{Graphical depiction of the registration an authentication process of fingerprint authentication.}~\label{fig:fingerprint}
\end{figure*}

\subsection{Smartphone-based authentication}

For this authentication scheme the user needs a smartphone with a camera to authenticate. It is a so-called \textit{token-based} authentication scheme. For the registration the user installs a smartphone app. The app generates and stores a secret, private key that is linked to the users' online account. 
When the authentication process on a PC is started, a coloured pattern similar to a QR code is generated. This code is scanned with the smartphone app. The secret key stored on the smartphone unlocks the pattern and reveals a PIN code. This PIN code is typed in at the PC to authenticate. 

The following figure \ref{fig:phototan} visualizes the registration (left) and authentication procedure (right). The registration process, in this case the installation of the app and the linking of the secret key with the online account, needs to be completed once. Afterwards, only the  authentication process takes place upon every log-in.

\begin{figure*}
  \centering
  \includegraphics[width=1\columnwidth]{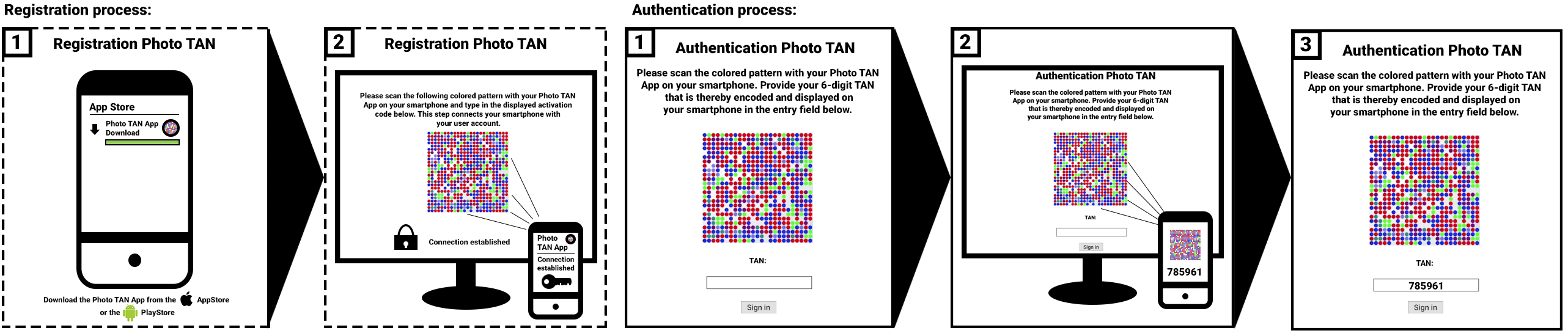}
  \caption{Graphical depiction of the registration an authentication process of PhotoTAN authentication.}~\label{fig:phototan}
\end{figure*}

\section{Scenario descriptions}

\subsection{Email Account}

Please imagine you'd like to check your inbox and send an email. You log into your mail account to gain access to your emails. 

\subsection{Social Network}

Please imagine you would like to log in to your social network to leave a post and to see your friends' news. 

\subsection{Online Banking}

Please imagine you aim to log in to your online banking account to check your balance and make a transfer of money. 

\subsection{Smart Home}

Please imagine several devices in your house or apartment are intelligently connected with each other. They react automatically to certain conditions or behaviours, e.g. your windows automatically close when rainfall is detected or your light automatically turns on when you enter a room. At 7am in the morning the coffee machine automatically heats up or you could check whether you have turned off your oven remotely from your office. To actively control the devices, or to change settings, you have to authenticate. 

\section{List of items - Context of use}

For every context of use the following questions were answered by the participants: 

\subsection{Familiarity with context}

\begin{itemize}
    \item Do you use [context of use and example)]?  - Yes/No
    \item If yes, what kind of [accounts/technologies depending on context] do you use? If you use several [accounts/technologies] please list all in the following text field. - text field
    \item If no, please try to imagine you were in the following scenario and try to respond to the questions accordingly: [scenario description]
    \item How worthy of protection do you deem the data protected in [context of use with examples for the data that would be concerned]? - continuous scale from 1 (not at all) to  100 (very much)
\end{itemize}

\subsection{Registration and authentication procedure of the authentication schemes}

The textual and visual description of each authentication scheme at the beginning of the study was followed by two items: 

\begin{itemize}
    \item How often do you use this authentication scheme? -  continuous scale ranging from 1 (not at all) to 100 (very often)
    \item How well do you know this authentication scheme (regardless of personal use)? -  continuous scale ranging from 1 (not at all) to 100 (very much)
\end{itemize}

\subsection{Preference rating}

\begin{figure*}
  \centering
  \includegraphics[width=1\columnwidth]{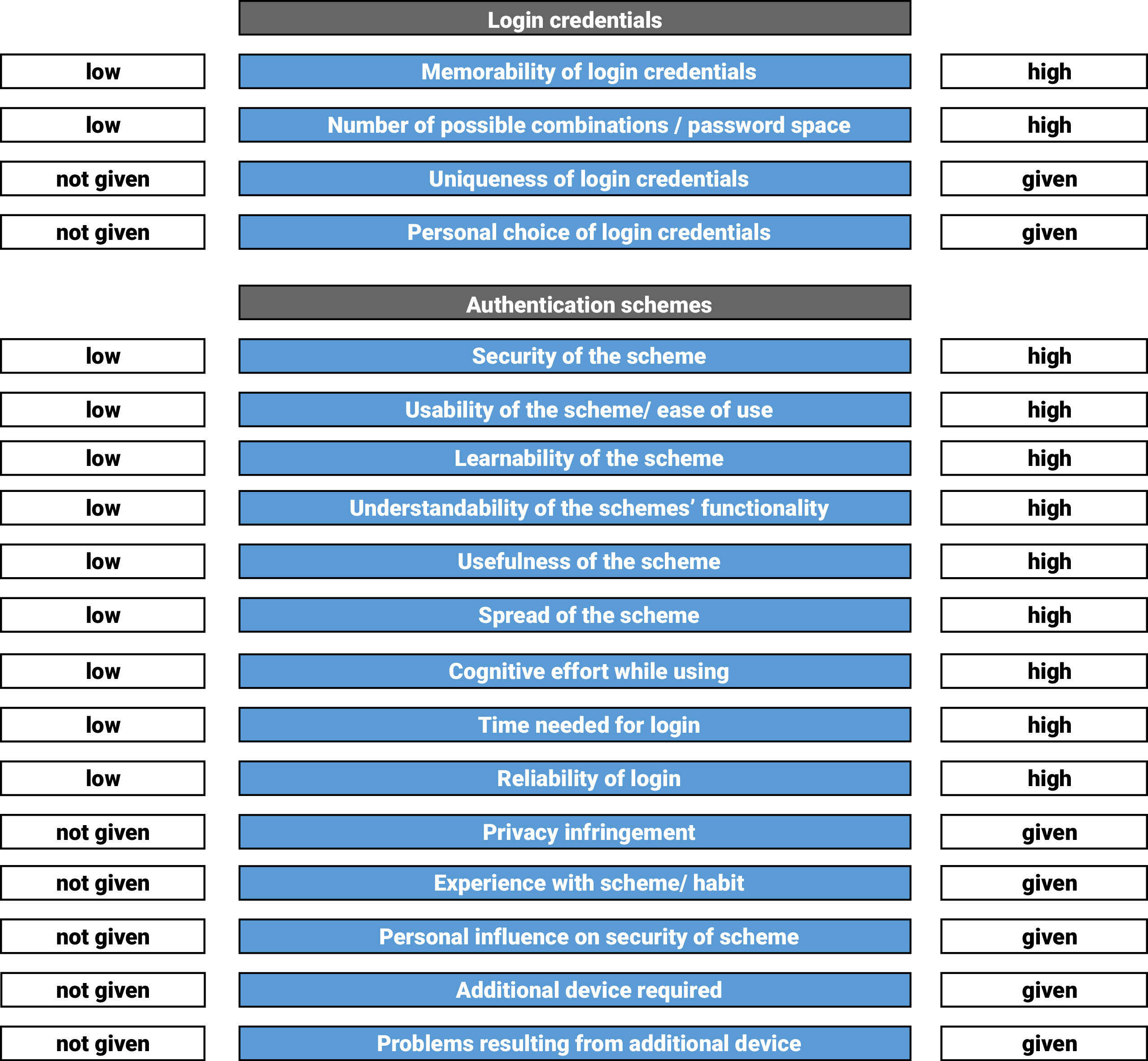}
  \caption{Graphical depiction of the matrix used to enable the selection of reasons for the preference rating.}~\label{fig:mcselection}
\end{figure*}

The participants were then asked to sequentially rate their preference for using each of the three authentication schemes in a certain context and to provide reasons for their preference rating in that certain context of use with the help of figure \ref{fig:mcselection}. 

\begin{itemize}
    \item Regardless of whether you already use or whether you would expect that authentication scheme in that context: How much would you like to use the following authentication scheme [name of scheme] for your [context of use]? - continuous scale from 1 (not at all) to 100 (very much)
    \item What were your reasons for the above preference rating of the [name of scheme] scheme? 
    
    To simplify the response, we have collected a selection of potential reasons for you.  Please only choose the reasons that have influenced your above decision and are applicable in terms of the above authentication scheme. 
    
    For every reason that played a role in your above decision please indicate how you rate the scheme by clicking on the field left or right of the criterion. 
    An example: If you like to use the above authentication scheme very much because you rate it easy to learn, please click ``high''  next to ``Learnability of the scheme''. Or if you do not like to use the scheme very much because you rate it insecure, please click ``low'' next to ``Security of the scheme''.
    If you do not find an applicable reason or like to add one, please state the reason net to ``other''. 
    Further explanations of the terms can be found here: [foldout button: Explanation of terms]
    
    \item Selection of reasons by clicking on buttons in Figure \ref{fig:mcselection}
\end{itemize}

\subsection{User perceptions of authentication schemes in the context of use}

\begin{itemize}
    \item \textit{Security:} As how secure do you perceive the authentication scheme for protecting your [context of use] data from unauthorized access or attacks? 
    - continuous scale from 1 (not at all) to 100 (very much)
     \item \textit{Privacy Concerns:} How strong are your concerns to provide the following data that the authentication scheme requires for logging in to your [context of use]? - password: your password, fingerprint: your fingerprint, Smartphone-based/PhotoTAN: your secret key to unlock the pattern stored on your smartphone - continuous scale from 1 (not at all strong) to 100 (very strong) 
    \item \textit{Effort-benefit ratio:} How do rate the effort-benefit ratio of the scheme for logging in to your [context of use]? Please consider the processes presented at the beginning of the study. - continuous scale from 1 (very bad) to 100 (very good)
    \item \textit{Intention to use:} How much would you intend to use the scheme for logging into your [context of use] if it was available to you? -  continuous scale from 1 (not at all) to 100 (very much)
\end{itemize}

\subsection{Alternative schemes}
\begin{itemize}
    \item Would you wish to use another authentication for this context of use that has not been presented in this study? If yes. which one? - text field
\end{itemize}

\section{List of items - overall evaluation}

Please finally rate the three authentication schemes against each other and independent of a certain context of use. 

\begin{itemize}
  \item \textit{Security:} As how secure do you perceive the authentication schemes in general for protecting your account from unauthorized access or attacks? - continuous scale from 1 (not at all) to 100 (very much)
     \item \textit{Privacy Concerns:} How strong are your concerns to provide the following data that the authentication scheme requires for logging in to your accounts in general? - password: your password, fingerprint: your fingerprint, Smartphone-based/PhotoTAN: your secret key to unlock the pattern stored on your smartphone - continuous scale from 1 (not at all strong) to 100 (very strong) 
    \item \textit{Effort-benefit ratio:} How do rate the effort-benefit ratio of the scheme for logging in to your accounts in general? Please consider the processes presented at the beginning of the study. - continuous scale from 1 (very bad) to 100 (very good)
    \item \textit{Efficiency:} How do you rate the efficiency of the schemes after some practice trials? - continuous scale from 1 (very bad) to 100 (very good)
    \item \textit{Usability:} Items of the System Usability Scale (SUS) by Brooke \cite{brooke1996sus} except for item 5 (integration of different functions)  - 5-point scale from 1 (completely disagree) to 5 (completely agree)
    \item \textit{Trust:} Items of the Facets of System Trustworthiness (FOST) Scale by Franke et al. \cite{franke2015advancing} in an adapted version for hypothesized trust. One additional item ``I think the authentication scheme is error-prone.'' -  6-point scale from 1 (completely disagree) to 6 (completely agree)
\end{itemize}

\section{Control variables and demographics}

\begin{itemize}
    \item Items of the Affinity for Technology Interaction (ATI) Scale by Franke et al. \cite{franke2019personal} - 6-point scale from 1 (completely disagree) to 6 (completely agree)
    \item Items of the scale for global information privacy concerns from Smith et al. \cite{smith1996information} and adapted by Malhotra et al. \cite{malhotra2004internet} - 7-point scale from 1 (strongly disagree) to 7 (strongly agree)
    \item gender - female, male, diverse/other
    \item age - in groups
    \item education - highest degree
    \item occupation - type of occupation
    \item IT (security) background - yes, no, type of IT (security) background
\end{itemize}

\section{H4: Reasons for authentication preference sorted by context of use}

\begin{figure*} [ht!]
  \centering
  \includegraphics[width=0.8\columnwidth]{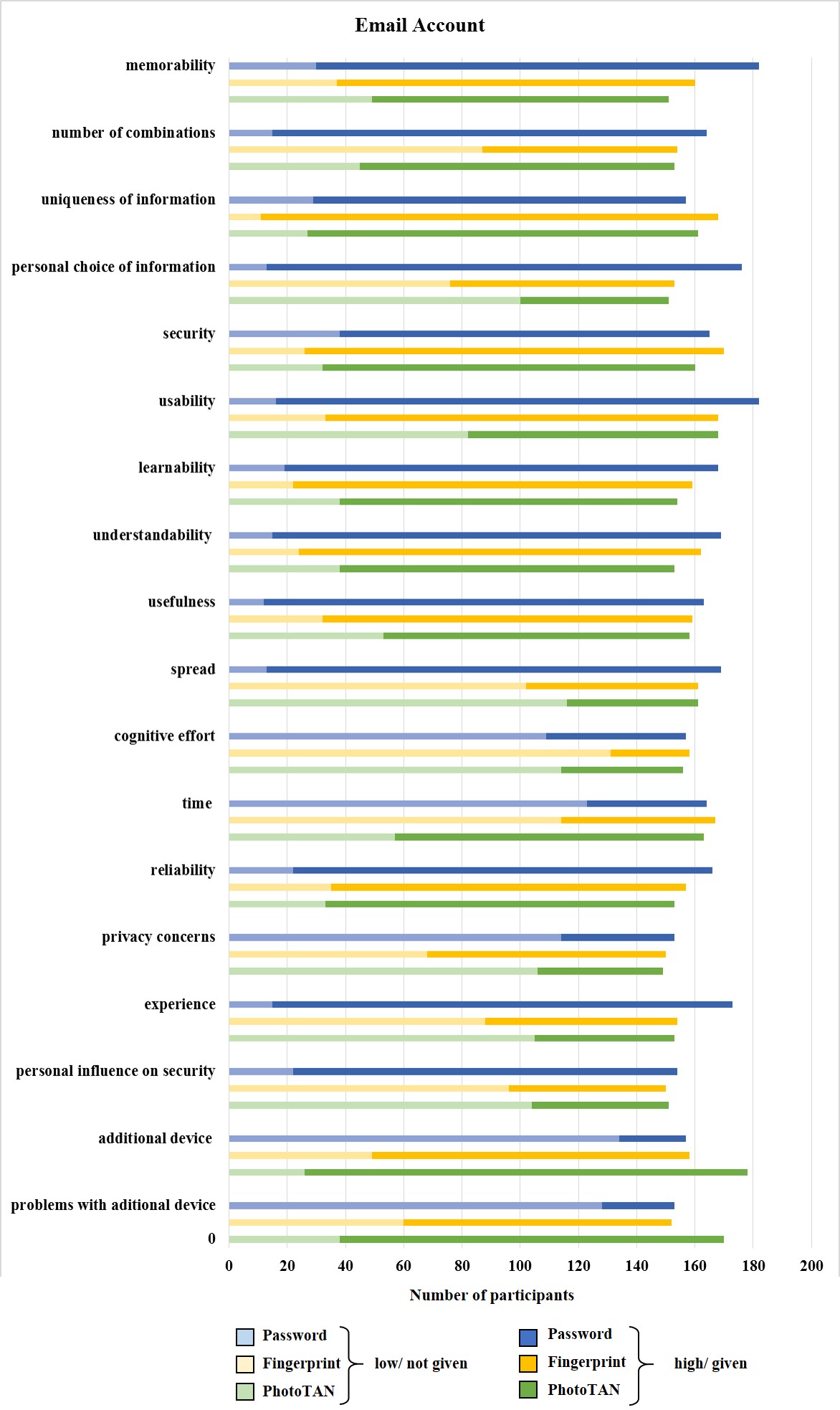}
  \caption{Graphical depiction of the qualitative feature rating related to the preference for an authentication scheme for the email account.}~\label{fig:mcemail}
\end{figure*}

\begin{figure*} [ht!]
  \centering
  \includegraphics[width=0.8\columnwidth]{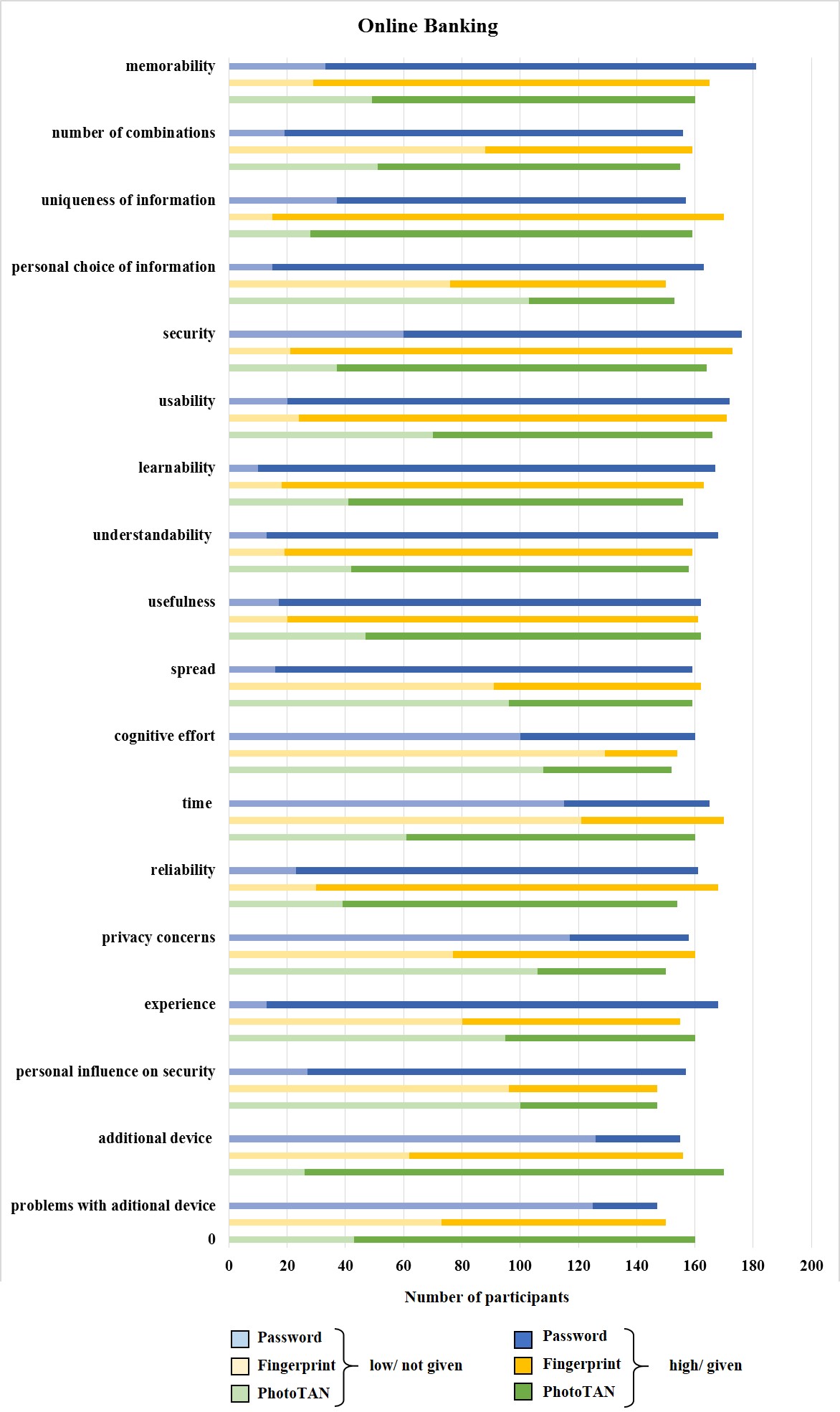}
  \caption{Graphical depiction of the qualitative feature rating related to the preference for an authentication scheme for the online banking account.}~\label{fig:mcob}
\end{figure*}

\begin{figure*} [ht!]
  \centering
  \includegraphics[width=0.8\columnwidth]{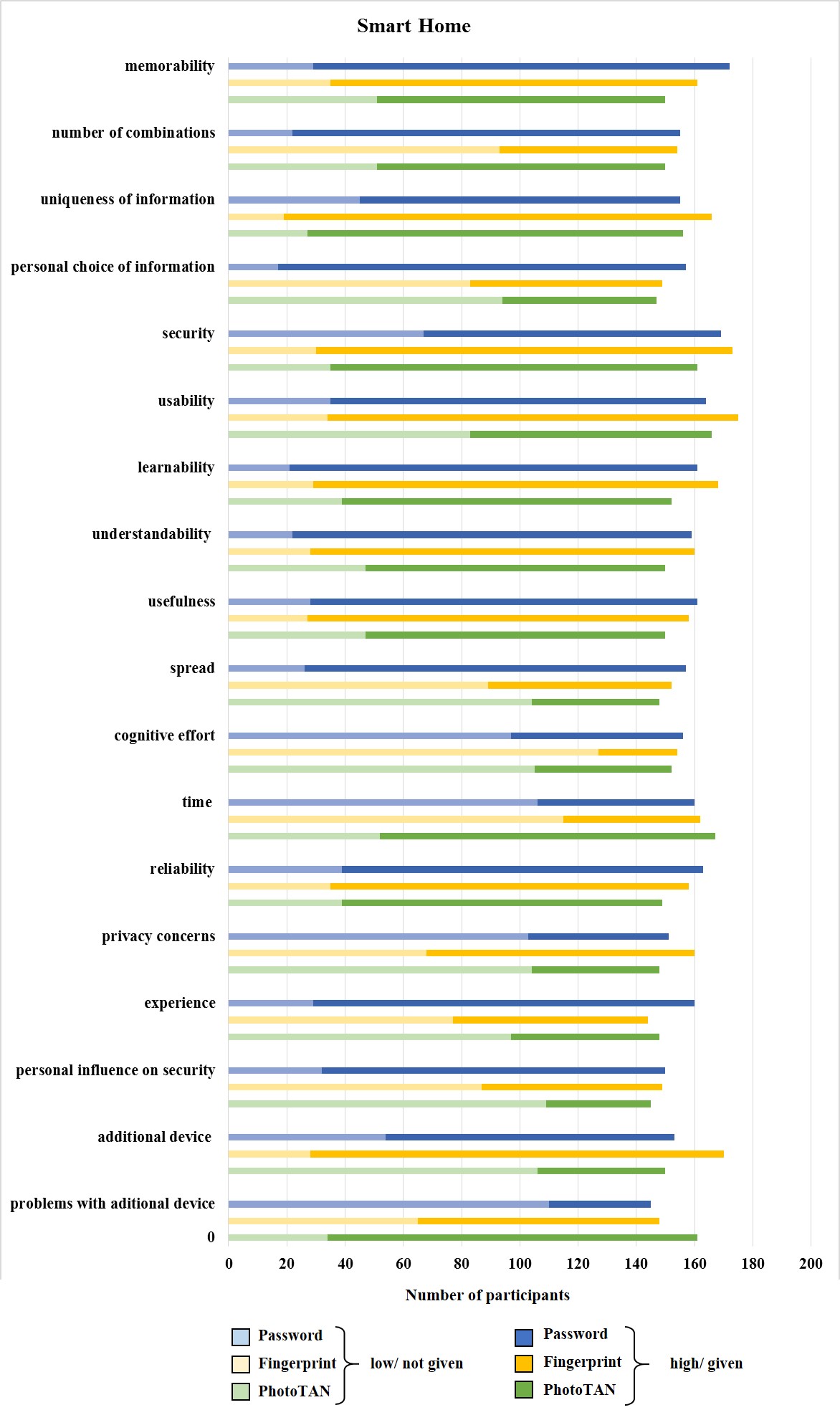}
  \caption{Graphical depiction of the qualitative feature rating related to the preference for an authentication scheme for the Smart Home.}~\label{fig:mcsh}
\end{figure*}

\begin{figure*} [ht!]
  \centering
  \includegraphics[width=0.8\columnwidth]{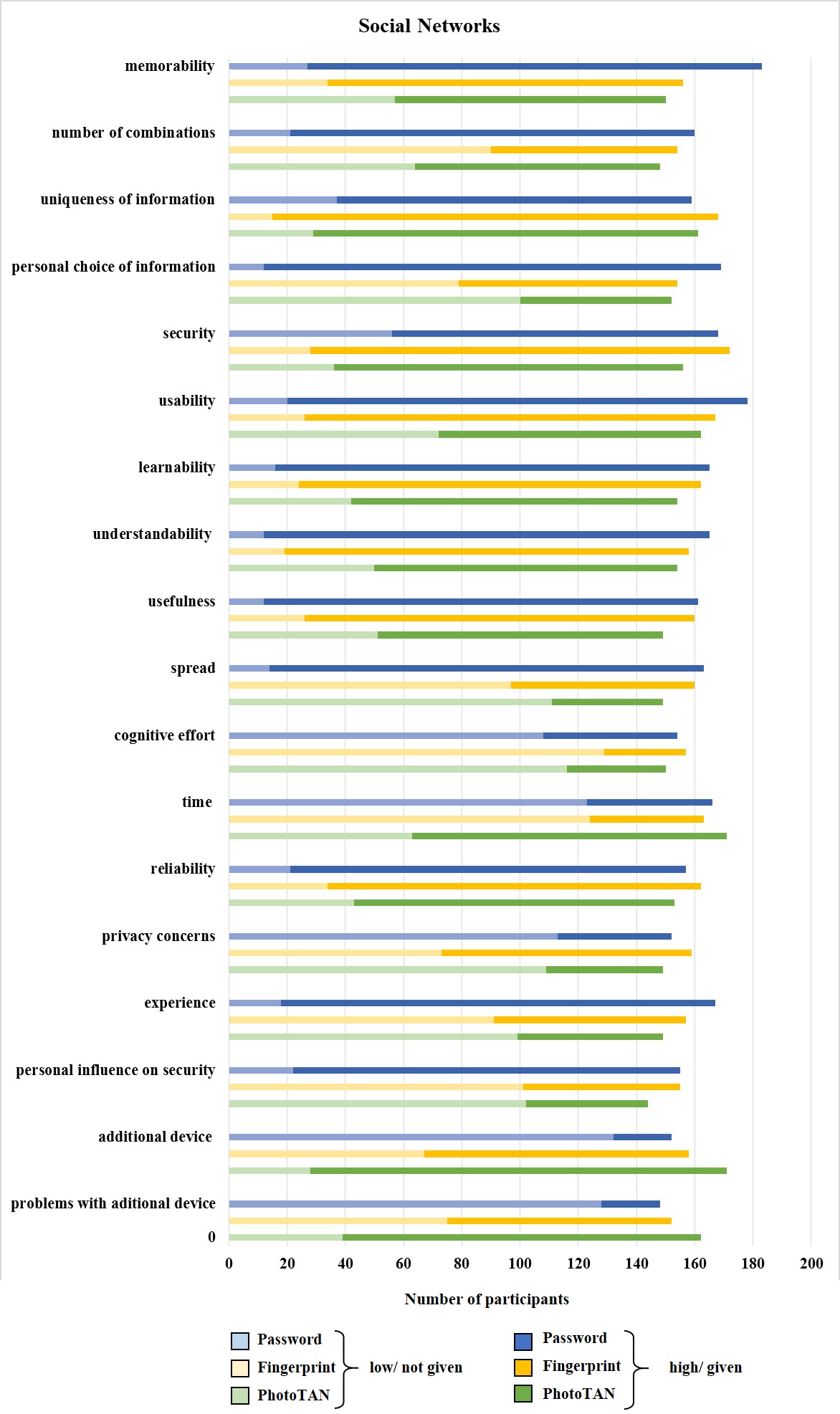}
  \caption{Graphical depiction of the qualitative feature rating related to the preference for an authentication scheme for the social network.}~\label{fig:mcsn}
\end{figure*}

The reasons for the user' preference ratings for authentication in a certain context of use, that were expressed in the selection of qualitative authentication scheme features, are depicted in Figure \ref{fig:mcemail}. Figure \ref{fig:mcob}. Figure \ref{fig:mcsh}. and Figure \ref{fig:mcsn}. 

\section{H4: Differences in reasons for authentication preference}

The following tables provide information on the frequency distribution of notions for the qualitative criteria for explaining the quantitative preference rating for each authentication scheme in each context of use. 

The top lines show the average number of qualitative criteria mentioned (maximum 18) and the absolute frequency of participants grouped by their quantitative preference rating. Overall, four preference categories were formed for participants showing a very low preference for using a certain scheme in the respective context with a score between 1 and 25 on a 100-point scale, participants with medium preference scores between 26 and 50, and 51 and 75, and finally participants with a very high preference and a score between 76 and 100. 

Below this, all 18 criteria are entered with their individual frequencies for each of the two possible selection options high versus low or given or not given, respectively. The first column of each preference category provides information on the relative proportion of persons in the respective preference category who specified the respective criterion as relevant for their quantitative preference rating, regardless of the option selected. This means that if 90 out of 100 participants in the lowest preference category (1-25 of 100 possible points) stated that the security of the procedure is relevant for their preference selection, 90.00 is entered in the third column of the table in the corresponding line. 

The colored arrows indicate how high the relative proportion of notions is in relation to all other notions for the same procedure in the same context in the same preference category. A red arrow pointing downwards indicates that it was mentioned as a criterion relatively seldom, a yellow one indicates a medium frequency and a green one a high relative frequency. 

In addition, the numbers in the second column of each preference category provide information on the absolute frequencies of the decisive proficiency. For example, if 50 of the 90 out of 100 participants in the lowest preference category were of the opinion that the level of certainty was low, the number 40  for ``high'' and the number 50 for ``low''  was entered in the fourth column of the corresponding row. The bars in the same row show the same ratio in visual form for easy interpretation. 

\includepdf[pages={2-13},scale=0.93,frame=false,angle=270]{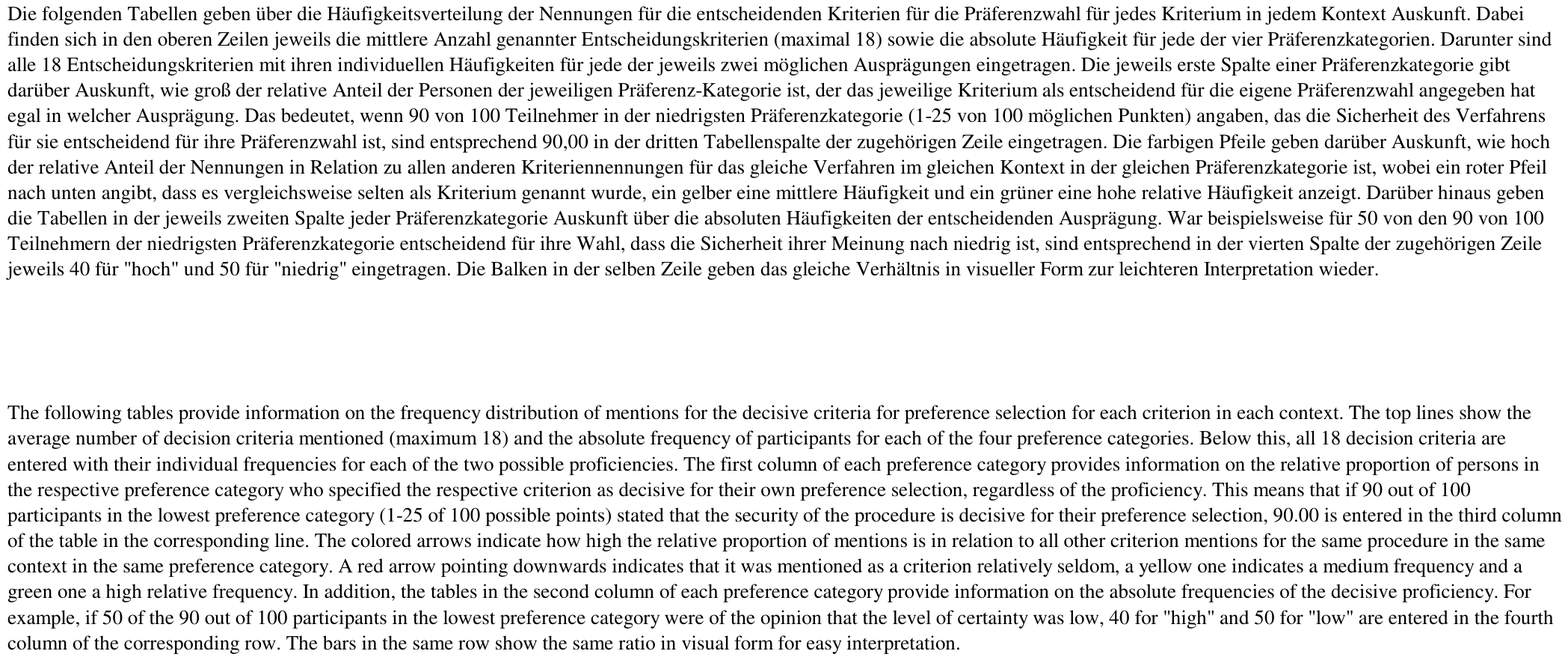}

\section{H4: Summary of the Multiple Regressions}

The following Tables 1, \ref{tab:MR_SN}, \ref{tab:MR_SH}, and
\ref{tab:MR_EM} show the results of the multiple regressions conducted for each context of use. Each table contains the standard and robust estimates for each predictor on the preference to use each authentication scheme in the respective context of use. The standard errors, the bias corrected confidence intervals, and the \textit{p}-values have been bootstrapped using 1000 samples.


\begin{table*}[ht!]
\caption{Results of the multiple regressions for the context online banking. Note: b = regression coefficient. SE = standard error.$\beta$ = standardized regression coefficient. p = bootstrapped p-value using 1000 samples. CI = 95\% bias corrected confidence interval. $b_r$ = robust regression coefficient. t = t-value.\label{tab:MR_SN1}}{
  \centering
\begin{tabular}{l | r r r r r r | r r r}
{\small{\textbf{Predictor}}} &	{\small{\textbf{b}}}&	{\small{\textbf{SE b}}}&	{\small{\textbf{$\beta$}}}&	{\small{\textbf{p}}}&	{\small{\textbf{Lower CI}}}&	{\small{\textbf{Upper CI}}}&	{\small{\textbf{$b_{r}$}}}&	{\small{\textbf{SE $b_r$}}}	&{\small{\textbf{t}}} \\
\hline
 \multicolumn{10}{c}{\small{\textbf{Preference for Password Authentication}}}\\
\hline
\small Constant&	16.17&	19.77&	/&	0.41&	-23.67&	54.06&	33.80&	20.34&	1.66 \\
\small Perceived Security&	0.56&	0.07&	0.55&	0.00&	0.41&	0.69&	0.65&	0.06&	11.64\\
\small Privacy Concerns&	-0.09&	0.05&	-0.10&	0.09&	-0.19&	0.00&	-0.11&	0.05&	-2.31\\
\small Perceived Effort-Benefit Ratio&	0.33&	0.09&	0.29&	0.00&	0.15&	0.50&	0.32&	0.07&	4.65\\
\small Protectworthiness of the data&	0.08&	0.09&	0.04&	0.40&	-0.11&	0.26&	0.05&	0.09&	0.53\\
\small General Efficiency of the scheme&	0.08&	0.10&	0.06&	0.47&	-0.11&	0.27&	0.02&	0.08&	0.26\\
\small SUS score for the scheme	& -0.12&	0.13&	-0.07&	0.36&	-0.41&	0.14&	-0.13&	0.12&	-1.07\\
\small Trust score for the scheme&	-1.01&	3.26&	-0.02&	0.77&	-6.77&	6.19&	-4.57&	2.86&	-1.60\\
\small Frequency of use of the scheme&	-0.11&	0.13&	-0.05&	0.41&	-0.41&	0.13&	-0.12&	0.13&	-0.90\\
\small Familiarity with the scheme&	0.11&	0.17&	0.04&	0.51&	-0.24&	0.43&	0.13&	0.16&	0.79\\
\hline
 \multicolumn{10}{c}{\small{\textbf{Preference for Fingerprint Authentication}}}\\
\hline
\small Constant&	16.39&	21.05&	/&	0.47&	-21.97&	58.15&	-10.42&	20.64&	-0.50\\
\small Perceived Security&	0.40&	0.09&	0.33&	0.00&	0.21&	0.56&	0.51&	0.09&	5.84\\
\small Privacy Concerns&	-0.11&	0.08&	-0.11&	0.12&	-0.28&	0.03&	-0.09&	0.06&	-1.36\\
\small Perceived Effort-Benefit Ratio&	0.23&	0.10&	0.19&	0.03&	0.02&	0.44&	0.23&	0.09&	2.68\\
\small Protectworthiness of the data&	-0.43&	0.10&	-0.22&	0.00&	-0.63&	-0.24&	-0.39&	0.12&	-3.33\\
\small General Efficiency of the scheme&	-0.08&	0.15&	-0.05&	0.58&	-0.36&	0.20&	-0.17&	0.13&	-1.27\\
\small SUS score for the scheme&	0.37&	0.15&	0.21&	0.02&	0.07&	0.69&	0.46&	0.14&	3.21\\
\small Trust score for the scheme&	6.17&	3.77&	0.10&	0.10&	-1.61&	13.08&	8.96&	3.98&	2.25\\
\small Frequency of use of the scheme&	0.07&	0.08&	0.06&	0.41&	-0.09&	0.21&	0.06&	0.07&	0.80\\
\small Familiarity with the scheme&	-0.08&	0.07&	-0.07&	0.23&	-0.21&	0.05&	-0.02&	0.07&	-0.31\\
\hline
 \multicolumn{10}{c}{\small{\textbf{Preference for PhotoTAN Authentication}}}\\
\hline
\small Constant&	-7.52&	19.26&	/&	0.69&	-41.03&	28.88&	-25.57&	17.68&	-1.45\\
\small Perceived Security&	0.39&	0.08&	0.35&	0.00&	0.22&	0.56&	0.49&	0.08&	6.17\\
\small Privacy Concerns&	-0.01&	0.08&	-0.01&	0.90&	-0.16&	0.14&	0.04&	0.07&	0.63\\
\small Perceived Effort-Benefit Ratio&	0.24&	0.09&	0.20&	0.02&	0.07&	0.41&	0.24&	0.09&	2.69\\
\small Protectworthiness of the data&	-0.26&	0.11&	-0.13&	0.03&	-0.49&	-0.05&	-0.13&	0.12&	-1.06\\
\small General Efficiency of the scheme&	-0.01&	0.09&	-0.01&	0.94&	-0.18&	0.15&	0.05&	0.10&	0.54\\
\small SUS score for the scheme&	0.34&	0.14&	0.20&	0.02&	0.07&	0.61&	0.34&	0.14&	2.33\\
\small Trust score for the scheme&	5.49&	3.64&	0.11&	0.14&	-2.39&	13.45&	4.50&	3.55&	1.27\\
\small Frequency of use of the scheme&	0.01&	0.09&	0.01&	0.91&	-0.18&	0.17&	0.02&	0.09&	0.21\\
\small Familiarity with the scheme&	0.01&	0.07&	0.01&	0.91&	-0.13&	0.17&	0.02&	0.07&	0.23\\
\hline
\end{tabular}}
    \end{table*}

\begin{table*}[ht!]
\caption{Results of the multiple regressions for the context social network. Note: b = regression coefficient. SE = standard error.$\beta$ = standardized regression coefficient. p = bootstrapped p-value using 1000 samples. CI = 95\% bias corrected confidence interval. $b_r$ = robust regression coefficient. t = t-value. \label{tab:MR_SN}}{
  \centering
\begin{tabular}{l | r r r r r r | r r r}
{\small{\textbf{Predictor}}} &	{\small{\textbf{b}}}&	{\small{\textbf{SE b}}}&	{\small{\textbf{$\beta$}}}&	{\small{\textbf{p}}}&	{\small{\textbf{Lower CI}}}&	{\small{\textbf{Upper CI}}}&	{\small{\textbf{$b_r$}}}&	{\small{\textbf{SE $b_r$}}}	&{\small{\textbf{t}}} \\
\hline
 \multicolumn{10}{c}{\small{\textbf{Preference for Password Authentication}}}\\
\hline
\small Constant	&	-2.83	&	21.74	&	/	&	0.90	&	-53.83	&	43.92	&	25.94	&	15.63	&	1.66	\\
\small Perceived Security	&	0.10	&	0.07	&	0.10	&	0.13	&	-0.01	&	0.22	&	0.02	&	0.05	&	0.31	\\
\small Privacy Concerns	&	-0.05	&	0.05	&	-0.10	&	0.27	&	-0.14	&	0.03	&	-0.03	&	0.04	&	-0.80	\\
\small Perceived Effort-Benefit Ratio	&	0.32	&	0.08	&	0.32	&	0.00	&	0.17	&	0.48	&	0.24	&	0.06	&	4.25	\\
\small Protectworthiness of the data	&	0.12	&	0.07	&	0.13	&	0.07	&	0.00	&	0.24	&	0.07	&	0.04	&	1.61	\\
\small General Efficiency of the scheme	&	0.09	&	0.11	&	0.07	&	0.38	&	-0.10	&	0.29	&	0.17	&	0.07	&	2.56	\\
\small SUS score for the scheme	&	0.22	&	0.11	&	0.17	&	0.06	&	0.00	&	0.44	&	0.23	&	0.09	&	2.49	\\
\small Trust score for the scheme	&	2.54	&	2.91	&	0.07	&	0.38	&	-3.64	&	8.99	&	1.97	&	2.35	&	0.84	\\
\small Frequency of use of the scheme	&	0.12	&	0.15	&	0.08	&	0.41	&	-0.20	&	0.47	&	0.26	&	0.11	&	2.49	\\
\small Familiarity with the scheme	&	0.00	&	0.24	&	0.00	&	1.00	&	-0.49	&	0.38	&	-0.30	&	0.13	&	-2.36	\\
\hline
 \multicolumn{10}{c}{\small{\textbf{Preference for Fingerprint Authentication}}}\\
\hline
\small Constant	&	-34.94	&	18.05	&	/	&	0.05	&	-73.16	&	-2.85	&	-29.28	&	20.12	&	-1.46	\\
\small Perceived Security	&	0.15	&	0.11	&	0.12	&	0.20	&	-0.08	&	0.39	&	0.21	&	0.10	&	2.15	\\
\small Privacy Concerns	&	-0.12	&	0.07	&	-0.11	&	0.12	&	-0.26	&	0.06	&	-0.09	&	0.07	&	-1.33	\\
\small Perceived Effort-Benefit Ratio	&	0.41	&	0.09	&	0.37	&	0.00	&	0.24	&	0.57	&	0.50	&	0.09	&	5.77	\\
\small Protectworthiness of the data	&	0.11	&	0.09	&	0.08	&	0.24	&	-0.07	&	0.27	&	0.11	&	0.09	&	1.32	\\
\small General Efficiency of the scheme	&	-0.06	&	0.14	&	-0.04	&	0.70	&	-0.32	&	0.18	&	-0.09	&	0.15	&	-0.61	\\
\small SUS score for the scheme	&	0.32	&	0.14	&	0.17	&	0.02	&	0.01	&	0.64	&	0.33	&	0.17	&	1.98	\\
\small Trust score for the scheme	&	6.51	&	4.44	&	0.10	&	0.15	&	-2.73	&	15.31	&	3.58	&	4.60	&	0.78	\\
\small Frequency of use of the scheme	&	0.12	&	0.08	&	0.10	&	0.11	&	-0.05	&	0.27	&	0.11	&	0.08	&	1.25	\\
\small Familiarity with the scheme	&	-0.07	&	0.08	&	-0.05	&	0.37	&	-0.21	&	0.09	&	-0.08	&	0.08	&	-0.97	\\
\hline
 \multicolumn{10}{c}{\small{\textbf{Preference for PhotoTAN Authentication}}}\\
\hline
\small Constant	&	-6.07	&	12.97	&	/	&	0.64	&	-30.68	&	20.59	&	-6.00	&	13.03	&	-0.46	\\
\small Perceived Security	&	0.21	&	0.07	&	0.20	&	0.00	&	0.06	&	0.35	&	0.19	&	0.08	&	2.37	\\
\small Privacy Concerns	&	-0.02	&	0.07	&	-0.02	&	0.76	&	-0.14	&	0.10	&	-0.02	&	0.07	&	-0.23	\\
\small Perceived Effort-Benefit Ratio	&	0.50	&	0.07	&	0.48	&	0.00	&	0.36	&	0.63	&	0.56	&	0.07	&	7.79	\\
\small Protectworthiness of the data	&	0.07	&	0.07	&	0.07	&	0.27	&	-0.06	&	0.20	&	0.07	&	0.07	&	0.95	\\
\small General Efficiency of the scheme	&	0.03	&	0.10	&	0.02	&	0.81	&	-0.16	&	0.19	&	0.04	&	0.10	&	0.38	\\
\small SUS score for the scheme	&	0.21	&	0.13	&	0.14	&	0.12	&	-0.03	&	0.47	&	0.15	&	0.13	&	1.12	\\
\small Trust score for the scheme	&	-3.29	&	3.20	&	-0.07	&	0.29	&	-10.76	&	3.13	&	-3.32	&	3.66	&	-0.91	\\
\small Frequency of use of the scheme	&	0.00	&	0.09	&	0.00	&	0.99	&	-0.18	&	0.17	&	0.03	&	0.09	&	0.31	\\
\small Familiarity with the scheme	&	0.00	&	0.08	&	0.00	&	0.97	&	-0.15	&	0.14	&	-0.01	&	0.07	&	-0.16	\\
\hline
\end{tabular}}
     \end{table*}

\begin{table*}[ht!]
\caption{Results of the multiple regressions for the context smart home. Note: b = regression coefficient. SE = standard error.$\beta$ = standardized regression coefficient. p = bootstrapped p-value using 1000 samples. CI = 95\% bias corrected confidence interval. $b_r$ = robust regression coefficient. t = t-value.\label{tab:MR_SH}}{
  \centering
\begin{tabular}{l | r r r r r r | r r r}
{\small{\textbf{Predictor}}} &	{\small{\textbf{b}}}&	{\small{\textbf{SE b}}}&	{\small{\textbf{$\beta$}}}&	{\small{\textbf{p}}}&	{\small{\textbf{Lower CI}}}&	{\small{\textbf{Upper CI}}}&	{\small{\textbf{$b_r$}}}&	{\small{\textbf{SE $b_r$}}}	&{\small{\textbf{t}}} \\
\hline
 \multicolumn{10}{c}{\small{\textbf{Preference for Password Authentication}}}\\
\hline
\small Constant	&	-1.03	&	25.18	&	/	&	0.98	&	-49.20	&	48.79	&	-7.98	&	24.94	&	-0.32	\\
\small Perceived Security	&	0.43	&	0.09	&	0.37	&	0.00	&	0.25	&	0.58	&	0.50	&	0.08	&	6.49	\\
\small Privacy Concerns	&	-0.02	&	0.07	&	-0.02	&	0.80	&	-0.18	&	0.12	&	-0.03	&	0.06	&	-0.45	\\
\small Perceived Effort-Benefit Ratio	&	0.35	&	0.09	&	0.32	&	0.00	&	0.16	&	0.53	&	0.37	&	0.08	&	4.80	\\
\small Protectworthiness of the data	&	0.22	&	0.07	&	0.21	&	0.00	&	0.07	&	0.38	&	0.20	&	0.07	&	3.10	\\
\small General Efficiency of the scheme	&	-0.11	&	0.11	&	-0.08	&	0.35	&	-0.34	&	0.12	&	-0.14	&	0.11	&	-1.35	\\
\small SUS score for the scheme	&	0.03	&	0.13	&	0.02	&	0.82	&	-0.23	&	0.27	&	0.02	&	0.15	&	0.13	\\
\small Trust score for the scheme	&	-0.71	&	3.71	&	-0.01	&	0.84	&	-7.91	&	7.55	&	1.58	&	3.79	&	0.42	\\
\small Frequency of use of the scheme	&	-0.05	&	0.16	&	-0.02	&	0.77	&	-0.36	&	0.27	&	-0.04	&	0.17	&	-0.24	\\
\small Familiarity with the scheme	&	0.12	&	0.21	&	0.04	&	0.51	&	-0.28	&	0.50	&	0.09	&	0.20	&	0.45	\\
\hline
 \multicolumn{10}{c}{\small{\textbf{Preference for Fingerprint Authentication}}}\\
\hline
\small Constant	&	-53.11	&	17.43	&	/	&	0.00	&	-87.71	&	-18.02	&	-57.96	&	14.34	&	-4.04	\\
\small Perceived Security	&	0.45	&	0.08	&	0.38	&	0.00	&	0.27	&	0.61	&	0.54	&	0.07	&	7.85	\\
\small Privacy Concerns	&	-0.10	&	0.06	&	-0.10	&	0.08	&	-0.22	&	0.00	&	-0.05	&	0.05	&	-1.05	\\
\small Perceived Effort-Benefit Ratio	&	0.39	&	0.09	&	0.31	&	0.00	&	0.21	&	0.58	&	0.41	&	0.06	&	6.47	\\
\small Protectworthiness of the data	&	0.08	&	0.07	&	0.06	&	0.27	&	-0.07	&	0.22	&	0.03	&	0.06	&	0.61	\\
\small General Efficiency of the scheme	&	-0.17	&	0.12	&	-0.11	&	0.16	&	-0.40	&	0.03	&	-0.14	&	0.10	&	-1.32	\\
\small SUS score for the scheme	&	0.29	&	0.14	&	0.16	&	0.04	&	0.01	&	0.59	&	0.23	&	0.11	&	2.07	\\
\small Trust score for the scheme	&	10.42	&	3.55	&	0.16	&	0.00	&	3.28	&	17.21	&	10.50	&	3.11	&	3.37	\\
\small Frequency of use of the scheme	&	0.07	&	0.07	&	0.06	&	0.35	&	-0.08	&	0.21	&	0.08	&	0.06	&	1.30	\\
\small Familiarity with the scheme	&	-0.01	&	0.06	&	-0.01	&	0.82	&	-0.13	&	0.12	&	0.02	&	0.06	&	0.37	\\
\hline
 \multicolumn{10}{c}{\small{\textbf{Preference for PhotoTAN Authentication}}}\\
\hline
\small Constant	&	1.44	&	13.82	&	/	&	0.91	&	-25.30	&	30.66	&	-1.55	&	13.83	&	-0.11	\\
\small Perceived Security	&	0.20	&	0.08	&	0.19	&	0.02	&	0.05	&	0.37	&	0.19	&	0.09	&	2.18	\\
\small Privacy Concerns	&	0.02	&	0.07	&	0.02	&	0.76	&	-0.12	&	0.16	&	0.04	&	0.07	&	0.55	\\
\small Perceived Effort-Benefit Ratio	&	0.34	&	0.08	&	0.32	&	0.00	&	0.15	&	0.51	&	0.42	&	0.09	&	4.91	\\
\small Protectworthiness of the data	&	0.03	&	0.07	&	0.03	&	0.67	&	-0.11	&	0.18	&	0.02	&	0.07	&	0.26	\\
\small General Efficiency of the scheme	&	-0.04	&	0.10	&	-0.04	&	0.67	&	-0.22	&	0.12	&	-0.04	&	0.10	&	-0.41	\\
\small SUS score for the scheme	&	0.40	&	0.14	&	0.26	&	0.01	&	0.08	&	0.71	&	0.42	&	0.15	&	2.84	\\
\small Trust score for the scheme	&	-3.71	&	3.74	&	-0.08	&	0.32	&	-11.23	&	2.74	&	-3.74	&	3.85	&	-0.97	\\
\small Frequency of use of the scheme	&	0.11	&	0.09	&	0.10	&	0.24	&	-0.08	&	0.30	&	0.14	&	0.09	&	1.44	\\
\small Familiarity with the scheme	&	-0.08	&	0.08	&	-0.09	&	0.34	&	-0.23	&	0.09	&	-0.10	&	0.07	&	-1.37	\\
\hline
\end{tabular}}
      \end{table*}

\begin{table*}[ht!]
\caption{Results of the multiple regressions for the context email account. Note: b = regression coefficient. SE = standard error.$\beta$ = standardized regression coefficient. p = bootstrapped p-value using 1000 samples. CI = 95\% bias corrected confidence interval. $b_r$ = robust regression coefficient. t = t-value. \label{tab:MR_EM}}{
  \centering
\begin{tabular}{l | r r r r r r | r r r}
{\small{\textbf{Predictor}}} &	{\small{\textbf{b}}}&	{\small{\textbf{SE b}}}&	{\small{\textbf{$\beta$}}}&	{\small{\textbf{p}}}&	{\small{\textbf{Lower CI}}}&	{\small{\textbf{Upper CI}}}&	{\small{\textbf{$b_r$}}}&	{\small{\textbf{SE $b_r$}}}	&{\small{\textbf{t}}} \\
\hline
 \multicolumn{10}{c}{\small{\textbf{Preference for Password Authentication}}}\\
\hline
\small Constant	&	11.37	&	20.27	&	/	&	0.57	&	-31.48	&	56.77	&	16.37	&	14.71	&	1.11	\\
\small Perceived Security	&	0.12	&	0.06	&	0.14	&	0.07	&	-0.02	&	0.25	&	0.02	&	0.05	&	0.35	\\
\small Privacy Concerns	&	-0.06	&	0.04	&	-0.09	&	0.16	&	-0.13	&	0.02	&	-0.02	&	0.04	&	-0.44	\\
\small Perceived Effort-Benefit Ratio	&	0.08	&	0.07	&	0.10	&	0.21	&	-0.04	&	0.23	&	0.00	&	0.05	&	0.02	\\
\small Protectworthiness of the data	&	0.22	&	0.07	&	0.24	&	0.00	&	0.07	&	0.35	&	0.23	&	0.05	&	4.57	\\
\small General Efficiency of the scheme	&	0.18	&	0.09	&	0.20	&	0.05	&	0.02	&	0.37	&	0.23	&	0.07	&	3.38	\\
\small SUS score for the scheme	&	0.25	&	0.11	&	0.19	&	0.02	&	0.04	&	0.44	&	0.31	&	0.09	&	3.54	\\
\small Trust score for the scheme	&	-0.04	&	2.43	&	0.00	&	0.99	&	-4.49	&	4.30	&	0.51	&	2.26	&	0.23	\\
\small Frequency of use of the scheme	&	-0.04	&	0.11	&	-0.03	&	0.69	&	-0.28	&	0.15	&	0.04	&	0.10	&	0.39	\\
\small Familiarity with the scheme	&	0.13	&	0.15	&	0.07	&	0.38	&	-0.12	&	0.42	&	0.02	&	0.12	&	0.14	\\
\hline
 \multicolumn{10}{c}{\small{\textbf{Preference for Fingerprint Authentication}}}\\
\hline
\small Constant	&	-52.59	&	17.02	&	/	&	0.00	&	-86.97	&	-19.44	&	-52.58	&	20.17	&	-2.61	\\
\small Perceived Security	&	0.24	&	0.08	&	0.19	&	0.00	&	0.06	&	0.39	&	0.25	&	0.10	&	2.61	\\
\small Privacy Concerns	&	-0.13	&	0.07	&	-0.12	&	0.07	&	-0.28	&	0.02	&	-0.14	&	0.07	&	-1.89	\\
\small Perceived Effort-Benefit Ratio	&	0.47	&	0.09	&	0.39	&	0.00	&	0.30	&	0.65	&	0.50	&	0.09	&	5.68	\\
\small Protectworthiness of the data	&	-0.02	&	0.09	&	-0.01	&	0.79	&	-0.21	&	0.14	&	-0.02	&	0.10	&	-0.16	\\
\small General Efficiency of the scheme	&	-0.15	&	0.12	&	-0.10	&	0.21	&	-0.41	&	0.06	&	-0.15	&	0.15	&	-1.04	\\
\small SUS score for the scheme	&	0.31	&	0.14	&	0.17	&	0.04	&	0.01	&	0.59	&	0.30	&	0.16	&	1.95	\\
\small Trust score for the scheme	&	10.35	&	3.51	&	0.17	&	0.00	&	3.39	&	17.85	&	10.25	&	4.37	&	2.35	\\
\small Frequency of use of the scheme	&	0.12	&	0.07	&	0.10	&	0.10	&	-0.03	&	0.26	&	0.12	&	0.08	&	1.56	\\
\small Familiarity with the scheme	&	0.08	&	0.07	&	0.07	&	0.31	&	-0.06	&	0.21	&	0.05	&	0.08	&	0.67	\\
\hline
 \multicolumn{10}{c}{\small{\textbf{Preference for PhotoTAN Authentication}}}\\
\hline
\small Constant	&	-13.62	&	15.66	&	/	&	0.39	&	-43.27	&	16.25	&	-18.89	&	15.29	&	-1.23	\\
\small Perceived Security	&	0.03	&	0.08	&	0.02	&	0.75	&	-0.13	&	0.19	&	0.00	&	0.08	&	0.04	\\
\small Privacy Concerns	&	0.02	&	0.08	&	0.02	&	0.78	&	-0.12	&	0.17	&	0.04	&	0.07	&	0.59	\\
\small Perceived Effort-Benefit Ratio	&	0.53	&	0.09	&	0.48	&	0.00	&	0.33	&	0.72	&	0.62	&	0.09	&	7.14	\\
\small Protectworthiness of the data	&	0.05	&	0.10	&	0.04	&	0.60	&	-0.14	&	0.26	&	0.06	&	0.09	&	0.67	\\
\small General Efficiency of the scheme	&	0.04	&	0.10	&	0.03	&	0.72	&	-0.15	&	0.22	&	0.03	&	0.10	&	0.26	\\
\small SUS score for the scheme	&	0.20	&	0.15	&	0.12	&	0.19	&	-0.10	&	0.51	&	0.17	&	0.15	&	1.14	\\
\small Trust score for the scheme	&	1.11	&	3.64	&	0.02	&	0.78	&	-7.05	&	8.67	&	1.91	&	3.79	&	0.50	\\
\small Frequency of use of the scheme	&	0.08	&	0.10	&	0.07	&	0.42	&	-0.13	&	0.28	&	0.11	&	0.09	&	1.14	\\
\small Familiarity with the scheme	&	-0.02	&	0.08	&	-0.02	&	0.81	&	-0.17	&	0.15	&	-0.04	&	0.07	&	-0.54	\\
\hline
\end{tabular}}
\end{table*}


\bibliographystyle{acm}
\bibliography{sample}